\documentclass[aps, twocolumn, prd, preprintnumbers, amsmath, amssymb, amsfonts, superscriptaddress, nofootinbib]{revtex4-1}
\pdfoutput=1

\usepackage[justification=raggedright,singlelinecheck=false]{caption}

\usepackage{amsmath,amssymb, bm, mathrsfs}
\usepackage{hyperref}
\usepackage{url}
\usepackage{graphicx}
\usepackage{color}
\usepackage{colortbl}
\usepackage{tabularx}
\usepackage{multirow}
\usepackage{comment}
\usepackage{slashed}
\usepackage{subcaption}
\usepackage{cancel}
\usepackage{braket}
\usepackage{soul}

\definecolor{Orange}{cmyk}{0,0.61,0.87,0}
\definecolor{JungleGreen}{cmyk}{0.99,0,0.52,0}
\definecolor{OliveGreen}{cmyk}{0.64,0,0.95,0.40}
\definecolor{Brown}{cmyk}{0,0.81,1,0.60}
\definecolor{RoyalBlue}{cmyk}{0.71,0.53,0,0.12}
\definecolor{Gray}{cmyk}{0,0,0,0.40}
\definecolor{LightPink}{cmyk}{0.0,0.25,0,0}
\definecolor{LLightPink}{cmyk}{0.0,0.10,0,0}
\definecolor{LightBlue}{cmyk}{0.25,0,0,0}
\definecolor{LightGray}{cmyk}{0,0,0,0.2}

\newcommand{\mr}[1]{\mathrm{#1}}

\newcommand{\Ht}[1]{\textcolor{Orange}{#1}}

\begin{document}
\hfill TU-1295

\title{Asymmetric Dark Matter from Low-Scale Spontaneous Leptogenesis} 

\author{Hiroki Takahashi }
\email{takahashi@hep-th.phys.s.u-tokyo.ac.jp}
\affiliation{Department of Physics, University of Tokyo, Bunkyo-ku, Tokyo
 113--0033, Japan}

\author{Juntaro Wada}
\email{juntaro.wada.e5@tohoku.ac.jp }
\affiliation{Department of Physics, Tohoku University, Sendai, Miyagi 980-8578, Japan}
\affiliation{Technical University of Munich (TUM), School of Natural Sciences, Physics Department, James-Franck-Str. 1, 85748 Garching, Germany,
}

\begin{abstract}
We investigate a novel type of asymmetric dark matter (ADM) model in which the dark matter asymmetry and the baryon asymmetry in our universe (BAU) are produced simultaneously via low-scale spontaneous leptogenesis, where the mass scale of right-handed neutrino is much lower than the Davidson–Ibarra bound $M_1 \ll 10^{9}~\mr{GeV}$. 
In our scenario, both asymmetries are predominantly sourced by a dynamical $CP$ phase, namely the majoron. 
Its kinetic misalignment provides a sufficiently large, time-dependent effective $CP$ phase, allowing efficient asymmetry production even for low-scale right-handed neutrinos. 
In our framework, the sources of $CP$ violation responsible for the BAU and ADM are correlated with each other, leading to a predictive relation for the dark matter mass. In particular, when the dark matter asymmetry reaches its equilibrium value before freeze-out, the dark matter mass is typically predicted to lie in the range
$\mathcal{O}(0.1)~\mathrm{GeV} \lesssim m_{\chi} \lesssim \mathcal{O}(100)~\mathrm{GeV}$,
which lies within the sensitivity of direct detection experiments.
On the other hand, if the dark matter asymmetry does not reach its equilibrium value due to weak coupling, the allowed mass range extends over a broader interval, 
$\mathcal{O}(0.1)~\mathrm{GeV} \lesssim m_{\chi} \lesssim \mathcal{O}(10)~\mr{TeV}$.

\end{abstract}

\maketitle

\section{Introduction}
The existence of dark matter (DM) and the observed baryon asymmetry of the universe (BAU) provide clear evidence for physics beyond the Standard Model (BSM). 
Cosmological observations indicate that the ratio of the present-day energy densities of DM and baryons is approximately given by $\Omega_{\mr{DM}} \simeq 5.4\,\Omega_B$~\cite{Planck:2018vyg}, which may suggest a close connection between their origins.  
Asymmetric dark matter (ADM)~\cite{Nussinov:1985xr, Kaplan:1991ah} provides an elegant framework that links the dark and baryonic abundances, positing that DM has a particle–antiparticle asymmetry analogous to that of ordinary matter.

ADM has been extensively investigated in various contexts, which can be classified into several categories.  
For instance, there are scenarios where the asymmetry is transferred from Standard Model (SM) sector to the dark sector~\cite{Hooper:2004dc, Kaplan:2009ag, Cohen:2009fz, Blennow:2010qp, Haba:2011uz, Servant:2013uwa}, those where the asymmetry is instead transferred from the dark sector to the SM~\cite{Foot:2003jt, Foot:2004pq, Shelton:2010ta, Haba:2010bm, Buckley:2010ui, Dutta:2010va}, and those where the asymmetries in both sectors are generated independently~\cite{Davoudiasl:2010am, Davoudiasl:2010am, Falkowski:2011xh, Feng:2013wn}.  
Among these, an especially interesting possibility is the scenario in which both the SM and DM are produced simultaneously through the decay of thermally generated right-handed neutrinos~\cite{Falkowski:2011xh, Falkowski:2017uya}.\footnote{
See also Refs.~\cite{Mojahed:2025exj, Abhishek:2026hex, McKenna:2026xar, Borah:2026vqw} for some recent work on ADM in seesaw models.
} 

Despite being a minimal and simple extension of thermal leptogenesis~\cite{Fukugita:1986hr}, this setup can account for the origins of both the baryonic and DM components of the universe. On the other hand, when the right-handed neutrino masses lie below the Davidson–Ibarra bound~\cite{Davidson:2002qv}, $M_1^{DI}\sim 10^{9}~\mr{GeV}$,\footnote{
More recently, it has been shown, based on a more comprehensive analysis including flavor effects~\cite{Nardi:2006fx, Abada:2006fw, Abada:2006ea, Dev:2017trv}, that $M_1 \gtrsim 10^9~\mr{GeV}$ is typically required~\cite{Granelli:2025cho}.
} this mechanism typically requires mass degeneracy or tuning of the CP phases~\cite{Pilaftsis:1997jf, Pilaftsis:1997dr, Pilaftsis:2003gt, Hambye:2003rt, Blanchet:2008pw}.

\begin{figure}[t]
  \centering
  {\includegraphics[width=0.45\textwidth]{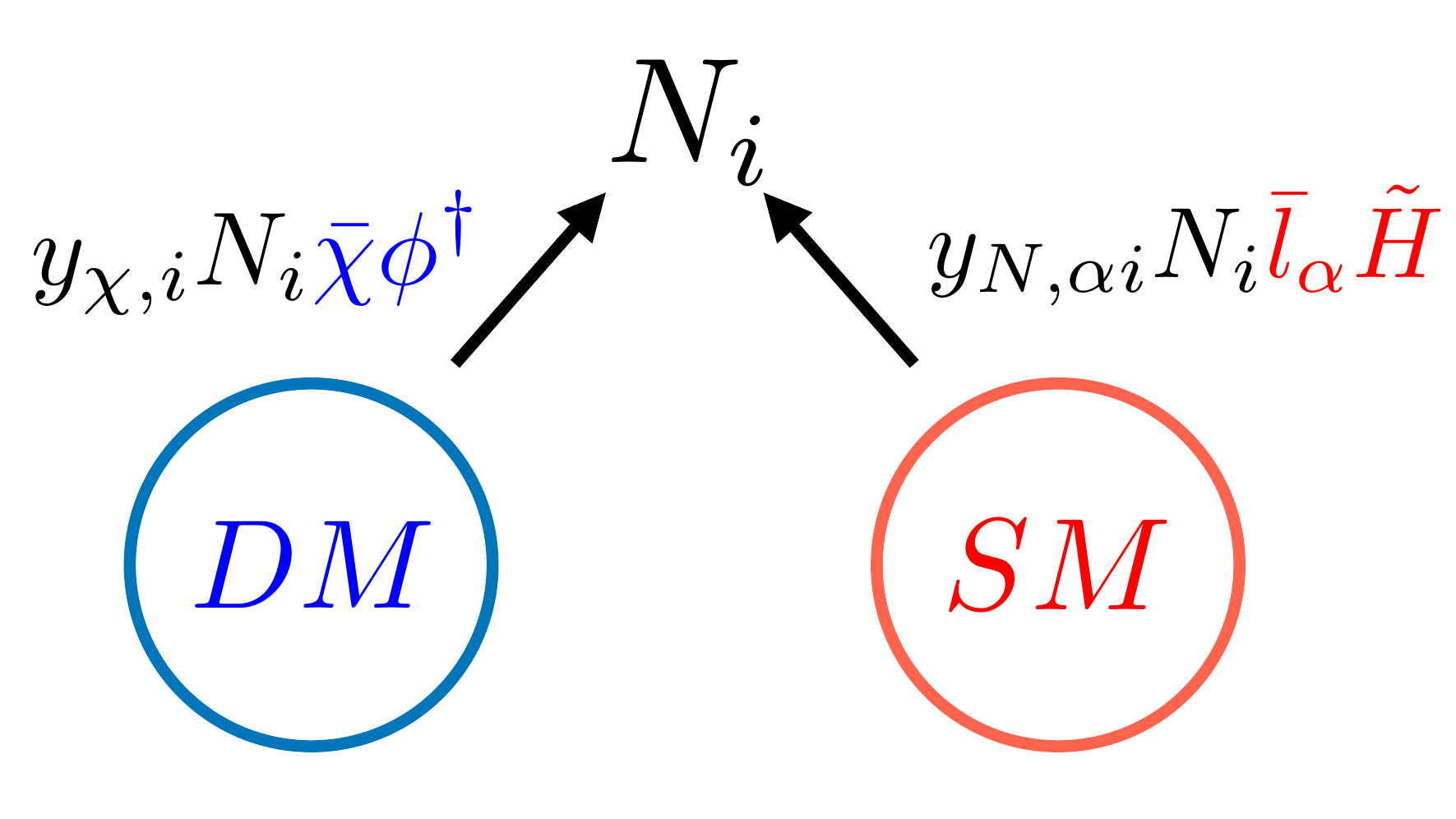}} 
  \caption{A schematic view of our framework:  the right-handed neutrinos are coupled to the SM leptons as well as to the DM particle through Yukawa interactions. In the early universe, when a dynamical $CP$ phase is present, the \emph{inverse} decay processes simultaneously generate SM lepton asymmetry and DM asymmetry. 
  }
  \label{fig: schematic view}
\end{figure}

In this work, we propose a new realization of ADM based on low-scale spontaneous leptogenesis. 
In contrast to conventional thermal leptogenesis, (low-scale) spontaneous leptogenesis~\cite{Chun:2023eqc, Barnes:2024jap, Wada:2024cbe}\footnote{
For previous (and recent) developments on spontaneous leptogenesis, see Ref.~\cite{Chiba:2003vp, Kusenko:2014lra, deCesare:2014dga, Pearce:2015nga, Kusenko:2014uta, Ibe:2015nfa, Bossingham:2017gtm, Domcke:2020kcp, Co:2020jtv, Domcke:2020quw, Berbig:2023uzs, Chao:2023ojl, Datta:2024xhg, Berbig:2025hlc, Peng:2025sri, Chun:2025abp}.} can achieve successful baryogenesis even for much lighter right-handed neutrinos than the Davidson–Ibarra bound~\cite{Davidson:2002qv}, if a sufficiently large dynamical $CP$ phase is provided through the kinetic misalignment mechanism~\cite{Co:2019jts, Chang:2019tvx}.  
We extend the model studied in previous work~\cite{Falkowski:2011xh, Falkowski:2017uya} by introducing a scalar field that spontaneously breaks the $U(1)_{B-L}$ symmetry, and investigate a spontaneous cogenesis scenario (See Refs.~\cite{March-Russell:2011ang, Kamada:2012ht} for pioneering studies of spontaneous cogenesis) where the majoron associated with the broken $U(1)_{B-L}$ acts as the background field.\footnote{
A similar mechanism appears in the Affleck–Dine cogenesis scenario~\cite{Cheung:2011if, vonHarling:2012yn, Borah:2022qln}, which also utilizes a time-dependent background field as a source of $CP$ violation.  
However, in that case, the asymmetries in the SM and DM are generated through the decay of the background field, whereas in spontaneous cogenesis, the background field merely acts as a coherent background without decaying.
}
We provide a schematic overview of our scenario in Fig.~\ref{fig: schematic view}.

As a consequence, the sources of $CP$ violation in the SM and ADM sectors, which were independent in the original setup~\cite{Falkowski:2011xh, Falkowski:2017uya}, become unified and are governed by a single dynamical $CP$ phase $\dot{\theta}$, originating from the kinetic motion of the majoron field.
Therefore, the baryon asymmetry $n_{\Delta B}$ and the DM asymmetry $n_{\Delta \chi}$ take the following form in thermal equilibrium:
\begin{align}
n_{\Delta B} &\propto \dot{\theta} T^2, \\
n_{\Delta \chi} &\propto \dot{\theta} T^2,
\end{align}
both being proportional to the same effective $CP$ phase $\dot{\theta}$.  
This property enables a predictive relation for the DM mass.  
Interestingly, depending on the strength of the Yukawa coupling between DM and the right-handed neutrino, the scenario can be categorized into two regimes: the \emph{freeze-out} scenario, in which the DM asymmetry reaches its equilibrium value before decoupling, and the \emph{freeze-in} scenario, in which the asymmetry freezes before reaching equilibrium.  
These two cases exhibit distinct behaviors in the predicted DM mass.

The remainder of this paper is organized as follows. In Section~\ref{sec: model}, we introduce our model, set our notation and conventions, and describe its basic properties. Section~\ref{sec: SLG} is devoted to a review of spontaneous leptogenesis. Section~\ref{sec: dark matter production} analyzes the generation of DM and baryon asymmetries, while Section~\ref{sec: discussion} presents phenomenological constraints on our scenario. We summarize our conclusions and discuss future prospects in Section~\ref{sec: summary}.

\section{Model}
\label{sec: model}

We extend the conventional type-I seesaw model~\cite{Minkowski:1977sc, Yanagida:1979as, Gell-Mann:1979vob, Mohapatra:1979ia} by introducing a global U(1)$_{B-L}$ symmetry and a dark gauged U(1)$_{D}$, which is already \emph{broken}, together with a dark sector that contains the DM particle: the interaction part of the Lagrangian is given by\footnote{
After the spontaneous breaking of the $U(1)_D$ symmetry, the $U(1)_D$-violating terms that can appear in the Lagrangian can be classified as follows: the mass term of the dark gauge boson, the mass term of the scalar field $\varphi_D$ responsible for the breaking of $U(1)_D$, and the Majorana mass term of the $U(1)_D$-charged fermion $\chi$, as well as its mixing with right-handed neutrinos, $ y_D \varphi_D \bar{\chi} N_i$.  
We assume that the scalar field $\varphi_D$ is sufficiently heavy, $m_{\varphi_D} \gg m_{Z'}, m_\chi$, and is completely decoupled from the other dark-sector particles.  
Moreover, if the charge of $\varphi_D$ satisfies $|q_{d, \varphi_D}| \geq 3$ and $|q_{d,\varphi_D}| \neq |q_{d,\chi}|$, then only the terms shown in the Lagrangian in Eq.~\eqref{Eq: Lagrangian} remain.  
This prescription for the charge assignment can alternatively be understood as introducing a $Z_2$ symmetry acting on $\chi$ and $\phi$, which results in the same Lagrangian~\eqref{Eq: Lagrangian}.
}

\begin{align}
\mathcal{L} 
    &=
        \mathcal{L}_{\text{SM}} + \mathcal{L}_{N} + \mathcal{L}_{\text{DS}}, \\
\mathcal{L}_{N}
    &= 
        {1 \over 2} \bar{N_i} i\slashed{\partial} N_i
        -\frac{1}{2} \sum_{i} g_{N,i} \Phi \bar{N}_{i}^c N_{i} \nonumber \\ 
    &\quad 
        -\sum_{\alpha,i} y_{N,\alpha i} \bar{l}_\alpha \tilde H N_i - \sum_{i} y_{\chi,i} \bar{\chi} \phi^\dagger N_i + \text{h.c.}, \\
\mathcal{L}_{\text{DS}}
    &=
        -{1\over 4} F^{\prime \mu \nu}F^\prime_{\mu \nu} 
    + {1 \over 2} m_{Z'}^2 Z^{\prime}_{\mu} Z^{\prime \mu}
    +\bar\chi(i\slashed{D} - m_\chi)\chi \nonumber \\
    &\quad 
        + |D \phi|^2  - V(\phi) - V(\Phi),
\label{Eq: Lagrangian}
\end{align}
where $\mathcal{L}_{\text{SM}}$ is the SM Lagrangian, and 
\begin{align}
\slashed{D} := \slashed{\partial} - i q_{d} g' Z^{\prime}_{\mu} \gamma^{\mu}
\end{align}
represents the covariant derivative associated with the $U(1)_{D}$ gauge interaction with the charge $q_d$. $\Phi$ denotes a complex scalar field with $B-L$ charge $+2$, $N_i$, $i=1,2,...$ are right-handed neutrinos with U(1)$_{B-L}$ charge $-1$.\footnote{More than one right-handed neutrinos are assumed to be introduced to make the model consistent with the neutrino oscillation data, but we do not specify the exact number in the following analysis, as in our scenario, the lightest species dominantly determines the asymmetries.} $l_\alpha,~\alpha =1,2,3$ and $\tilde{H} \equiv i\sigma_2 H^*$ are the SM lepton and Higgs doublets, respectively, while $\chi$ and $\phi$ represent dark sector fermion and complex scalar with $B-L$ charge $-1$ and $0$, respectively. $F^{\prime \mu \nu}$ is field strengths of the dark gauge field $Z^{\prime}_{\mu}$. 

We assume that $\chi$ and $\phi$ have masses $m_\chi$ and $m_\phi$ with $m_\chi < m_\phi$.
The $U(1)_D$ gauge interaction ensures that the symmetric component of $\phi$ and $\chi$ efficiently annihilates away, and we assume that $m_{Z'} < m_{\chi}$, as we will discuss later.
To ensure the stability of $\chi$, we also impose a $\mathbb{Z}_2$ symmetry, under which $\phi$ and $\chi$ are charged while the other particles are not charged.

We assume that the mixing between $\phi$ and $\Phi$ is negligibly small so that their potentials are separable. Although we do not specify the potential for $\phi$, it is assumed that it does not develop a VEV.
On the other hand, the potential for $\Phi$, $V(\Phi)$, is assumed to be such that $\Phi$ develops a VEV, which breaks the $B-L$ symmetry. We parameterize $\Phi$ after the $B-L$ symmetry breaking as follows:
\begin{align}
    \Phi = \frac{f}{\sqrt{2}} e^{iJ/f},
\end{align}
where $f$ denotes the decay constant and $J$ is the pseudo Nambu-Goldstone mode called majoron. The breaking of $B-L$ gives majorana masses to the right-handed neutrinos $M_i=fg_{N,i}/\sqrt{2}$.

From the perspective that global symmetries are necessarily broken by quantum gravity~\cite{Banks:2010zn, Witten:2017hdv, Harlow:2018jwu, Harlow:2018tng}, it is natural to expect the existence of higher-dimensional operators that break global $U(1)_{B-L}$ symmetry, as introduced in
\begin{align}
\label{eq: higher dimensional operator}
V(\Phi)_{\cancel{B-L}}=c_n {\Phi^n \over M_{\mr{pl}}^{n-4}} + \rm{h.c.},
\end{align}
where $M_{\mr{pl}}$ is the reduced Planck scale, and $c_n$ with $n>4$ is a numerical constant. This higher-dimensional operator plays an especially important role in the early universe as a source of $B-L$ violation~\cite{Berbig:2023uzs, Wada:2024cbe}, and it provides the origin of the majoron mass.

For later convenience, we perform the field redefinition\footnote{We note that this field redefinition may flip its sign depending on the $B-L$ charge assignment of $\Phi$.  
For instance, when $q_{\Phi} = -2$, the Majorana mass term can be written as
\begin{align}
\mathcal{L}_{N} 
	\supset -\frac{1}{2} \sum_{i} g_{N,i}\, \Phi\, \bar{N}_{i} N_{i}^{c} ,
\end{align}
which corresponds to swapping $\Phi$ and $\Phi^{*}$ in the original Eq.~\eqref{Eq: Lagrangian}. In this case, we keep $q_{N_i} = -1$, $q_{\chi} = -1$, and $q_{\psi}$ unchanged, but the field redefinition must be performed in the opposite direction compared to the case with $q_{\Phi} = 2$.  
Nevertheless, Eqs.~\eqref{eq: redefinition N}, \eqref{eq: redefinition chi}, and \eqref{eq: redefinition psi} still provide the correct redefinitions, if one instead takes $q_{\Phi} = -2$.
} 
\begin{align}
\label{eq: redefinition N}
N_i &\to e^{i q_{N_i} (q_{\Phi} / 2) \theta / 2}\, N_i, \\
\label{eq: redefinition chi}
\chi &\to e^{i q_{\chi} (q_{\Phi} / 2) \theta / 2}\, \chi, \\
\label{eq: redefinition psi}
\psi &\to e^{i q_{\psi} (q_{\Phi} / 2) \theta / 2}\, \psi,
\end{align}
where $\psi$ denotes SM fermion, $\theta \equiv J/f$, and $q_{\cdots}$ denotes $B-L$ charge of the each field; $q_{\Phi} = 2$, $q_{N_i} = -1$, $q_{\chi} = -1$ respectively. These redefinitions transform the coupling between the right-handed neutrinos and the majoron into the derivative coupling:
\begin{align}
\label{eq: induced term from majoron back ground}
\Delta \mathcal{L}
	&= -\,{q_{\Phi} \over 4} \partial_\mu \theta J^\mu_{B-L}\\
    &:=
		-\,{q_{\Phi} \over 4} (\partial_{\mu}\theta) \Bigl( q_{N_i}\, \, \bar{N}_i\gamma^{\mu} N_i
		+\, J^{\mu}_{\chi,\,B-L}
            \nonumber \\
    &\qquad 
		+\, \sum_{\psi \in \mathrm{SM}} J^{\mu}_{\psi,\,B-L} \Bigr),
\end{align}
where $J^{\mu}_{\chi,\,B-L} := q_{\chi}\, \bar{\chi}\gamma^{\mu}\chi$ and $\sum_{\psi \in \mathrm{SM}} J^{\mu}_{\psi,\,B-L} := \sum_{\psi \in \mathrm{SM}} q_{\psi}\, \bar{\psi}\gamma^{\mu}\psi$  represent the $B-L$ currents of the DM particle $\chi$ and the SM fermions. 
This derivative coupling, in a non-trivial $\dot{\theta}$ background, induces the splitting of energy levels between particles and anti-particles~\cite{Ibe:2015nfa, Chun:2023eqc} and leads to a scenario so-called spontaneous leptogenesis, which we will discuss in Sec~\ref{sec: SLG}.

Furthermore, we limit ourselves to the case of hierarchical right-handed neutrino masses, $M_1 \ll M_{i}$, $i = 2,...$ 
such that only the lightest right-handed neutrino contributes significantly to the generation of asymmetries.\footnote{If their masses are too hierarchical, however, heavier elements might induce the standard thermal leptogenesis contribution. We will not consider such a complication.}

\section{Low-scale spontaneous leptogenesis}
\label{sec: SLG}

In this section, we review low-scale spontaneous leptogenesis in the context previously discussed in Refs.~\cite{Chun:2023eqc, Barnes:2024jap, Wada:2024cbe}, where the scale of the leptogenesis is much lower than the Davidson–Ibarra bound $M_1 \ll 10^{9}~\mr{GeV}$~\cite{Davidson:2002qv}.
The key ingredient of spontaneous leptogenesis is a nonzero background of the majoron field, $\dot{\theta}$, which turns the derivative coupling $\partial_\mu \theta J^\mu_{B-L}$ into a CPT-violating term $\dot{\theta} J^0_{B-L}$~\cite{Chiba:2003vp, Kusenko:2014uta, Ibe:2015nfa}. 
This term can be shown to induce level splitting among particles and anti-particles~\cite{Cohen:1987vi, Cohen:1988kt},which eventually leads to the source term of $B-L$ asymmetry in the Boltzmann equation whenever a $B-L$ violating interaction is in equilibrium~\cite{Ibe:2015nfa, Chun:2023eqc, Wada:2024cbe}.

We stress that in the conventional thermal leptogenesis scenario with right-handed neutrinos at a low scale, the baryon asymmetry of the universe cannot be explained unless the CP violation 
is enhanced through mass degeneracy or tuning of the CP phases~\cite{Pilaftsis:1997jf, Pilaftsis:1997dr, Pilaftsis:2003gt, Hambye:2003rt, Blanchet:2008pw}. 
However, in spontaneous leptogenesis, if a sufficiently large dynamical CP phase background, 
$\dot{\theta}$ exists, low-scale leptogenesis can be realized without these parameter tunings, and such a background can be achieved through a kinetic misalignment scenario~\cite{Co:2019jts, Chang:2019tvx}.\footnote{
When $\dot{\theta}$ originates from the misalignment of the majoron field, a much heavier right-handed neutrino mass, $M_1 > 10^{10}~\mr{GeV}$, is required~\cite{Ibe:2015nfa, Chun:2023eqc}. This does not correspond to the situation of our interest.
}
It should also be stressed that this mechanism generates baryon asymmetry without requiring all of Sakharov’s conditions to be satisfied, because of the dynamical violation of CPT invariance.

Assuming that the contribution from the lightest right-handed neutrino dominates, the Boltzmann equation for the lepton asymmetry density $n_{\Delta l_\alpha}:= n_{l_\alpha}-n_{\bar{l}_\alpha}$ in a non-zero $\dot{\theta}$ background is given by the wash-in type equation~\cite{Domcke:2020quw, Chun:2023eqc, Wada:2024cbe}:
\begin{align}
\label{eq: Boltzmann equation for lepton sector}
\dot{n}_{\Delta l_{\alpha}} + 3 H n_{\Delta l_{\alpha}} = - n_{N_1}^{\rm{eq}}\langle{\Gamma_{N_1 \to l_{\alpha} H}\rangle} \left(\frac{n_{\Delta l_\alpha}}{n_{l_\alpha}^{\rm{eq}}} +\frac{n_{\Delta H}}{n_H^{\rm{eq}}} -\frac{\dot{\theta}}{T}\right),
\end{align}
where $n_{\Delta H} := n_{H} - n_{\bar{H}}$ is the Higgs asymmetry density, $n_{X}^{\mr{eq}}, X=l_{\alpha},N_1, H$ represent the equilibrium number density of $X$, and $\langle{\Gamma_{N_1 \to l_{\alpha} H}\rangle}$ is the thermally-averaged decay width given by
\begin{align}
\label{eq: thermal averaged interaction rate}
\langle{\Gamma_{N_1 \to l_{\alpha} H} \rangle}:= {K_1(z)\over K_2(z)}\Gamma_{N_1 \to l_{\alpha} H},
\end{align}
with $K_{1}$ and $K_{2}$ being modified Bessel functions of the first and second kind, respectively, $\Gamma_{N_1 \to l_{\alpha} H}= (|y_{N,\alpha1}|^2/16\pi)M_1$,
and $z:=M_1/T$. 
Here, we ignore CP-violating decays, which is a good approximation when the mass scale of the right-handed neutrino is far below
the Davidson–Ibarra bound, $M_1 \ll 10^{9}~\mathrm{GeV}$~\cite{Davidson:2002qv}.

In addition to decays and inverse decays, there also exist processes that violate only the SM lepton number, such as $\Delta L = 1$ scattering~\cite{Pilaftsis:2003gt, Giudice:2003jh} and $\Delta L = 2$ scattering processes~\cite{Giudice:2003jh}. These processes may also contribute to the lepton asymmetry through the majoron background. 
In particular, the $\Delta L = 1$ scatterings scale as $|y_{N,\alpha1}|^2$, similarly to inverse decays, but their impact on the final asymmetry is a subleading effect. This is because the regime in which $\Delta L = 1$ scatterings contribute as the leading effect to the asymmetry is restricted to the high-temperature region $z< 1$~\cite{Buchmuller:2004nz, Hahn-Woernle:2009jyb}. In our setup, however, the final asymmetry is determined at $z$ larger than $1$. (See the paragraph including Eq.~\eqref{eq: SM equilibrium range}.)\footnote{
\label{note: Delta 1 scattering}
Conversely, $\Delta L = 1$ scatterings can have a non-negligible impact on the asymmetry in the weak-washout regime, $K \lesssim 1$ with $K$ denoting the washout parameter~\cite{Buchmuller:2004nz} (See also the paragraph including Eq.~\eqref{eq: not-hierarchal coupling}), where washout effects cannot fully erase the asymmetry generated at $z<1$~\cite{Buchmuller:2004nz, Hahn-Woernle:2009jyb}. 
}

The $\Delta L = 2$ processes that violate only the SM lepton number can be neglected since their rates are proportional to $|y_{N,\alpha1}|^4$ and thus higher order in $|y_{N,\alpha1}| \ll 1$. There also exist $\Delta L = 2$ scattering processes involving dark matter, such as $\chi\phi \leftrightarrow \bar{L}\,\bar{H}$. We do not consider these processes either. The validity of this assumption will be discussed in Sec.~\ref{sec: dark matter production}.

Several important features of spontaneous leptogenesis are understood from the Boltzmann equation. 
Firstly, the terms in Eq.~(\ref{eq: Boltzmann equation for lepton sector}) correspond to the contribution from the inverse decay, which, in conventional leptogenesis models, acts as a wash-out term. Moreover, unlike thermal leptogenesis, low-scale spontaneous leptogenesis takes place when the right-handed neutrinos are in thermal equilibrium. 
If the inverse decay is efficient, from the Boltzmann equation, the lepton asymmetry is estimated to be~\cite{Chun:2023eqc} 
\begin{align}
\label{eq: lepton asymmetry in terms of number density}
&n_{\Delta L} = \frac{c_{L}}{6} \dot{\theta} T^2, 
\end{align}
where $c_{L}$ is the coefficient determined by the chemical equilibrium conditions.
For example, at $T < 10^5~\mr{GeV}$, $c_L \simeq 51/26$. 

The baryon asymmetry can be produced from this lepton asymmetry through the sphaleron processes~\cite{Kuzmin:1985mm}. If the inverse decay was decoupled before the sphaleron process was, we obtain~\cite{Chun:2023eqc} 
\begin{align}
\label{eq: baryon asymmetry in terms of number density}
&n_{\Delta B} = \frac{c_{B}}{6} \dot{\theta} T^2, 
\end{align}
where $c_{B}$ is the coefficient determined by the chemical equilibrium conditions. For example, at $T < 10^5~\mr{GeV}$, $c_B = -14/13$.
For details of the Boltzmann equation used in the numerical computation as well as the derivation of the equilibrium values of the asymmetries, Eqs.~(\ref{eq: lepton asymmetry in terms of number density}) and (\ref{eq: baryon asymmetry in terms of number density}), see Appendix~\ref{appendix: Boltzmann equation}.

In our scenario, leptogenesis proceeds predominantly through inverse decays, which operate efficiently only in thermal equilibrium. This requires the so-called strong washout condition, $\sum_{\alpha }\Gamma_{N_1}(N_1 \to l_{\alpha} H) > H_1$ where $H_1 = H(T=M_1)$ and $H(T)$ denotes the Hubble parameter at the radiation dominant epoch. 
It is customary to express this condition in terms of the decay parameter, $K:=\tilde{m}_1/m_{*} >1$~\cite{Buchmuller:2004nz}, where $\tilde{m}_1:=(\sum_{\alpha} |y_{N,\alpha 1}|^2) v^2/2M_1$ is the effective neutrino mass, and $m_{*}\simeq 1\times 10^{-3}~\mr{eV}$ denotes the equilibrium neutrino mass. 
For an effective mass of order the atmospheric scale, i.e., $\tilde{m}_1 \simeq O(0.05)~\mr{eV}$, one finds $K \sim 50$, confirming that the strong washout condition is indeed satisfied. Hereafter, we assume that flavor structure in yukawa coupling $y_{N, \alpha 1}$ is not hierarchal for simplicity:
\begin{align}
\label{eq: not-hierarchal coupling}
|y_{N, e 1}|^2 \simeq |y_{N, \mu 1}|^2 \simeq |y_{N, \tau 1}|^2,
\end{align}
and we define $y_{N,1} := \sqrt{(\sum_{\alpha} |y_{N,\alpha1}|^2)}$ as a typical Yukawa coupling to the SM lepton and Higgs.

To evaluate the resulting baryon asymmetry, it is essential to determine the temperature range in which inverse decays remain in equilibrium. 
For this purpose, it is convenient to introduce the conventional function~\cite{Buchmuller:2004nz},
\begin{align}
\label{eq: inverse decay criterion for leptogenesis}
W_{\rm{ID}}^{L}(z)
:= z \frac{\sum_{\alpha}\langle{\Gamma_{N_1 \to l_\alpha H}\rangle}}{H_1}\frac{n^{\rm{eq}}_{N_1}}{n^{\rm{eq}}_{l_{\alpha}}}.
\end{align}
The $W_{\rm{ID}}^L(z)$ quantifies the efficiency of inverse right-handed neutrino decays involving the SM particle; if $z W_{\rm{ID}}^{L}(z) > 1$,\footnote{In Ref.~\cite{Buchmuller:2004nz}, the criterion $W_{\rm ID}^{L}(z) > 1$ was introduced to determine whether the washout processes are in equilibrium. However, the condition that is actually equivalent to the equilibrium requirement $\langle \Gamma_{N_1 \to l_{\alpha} H} \rangle > H(T)$ is rather $z\, W_{\rm ID}^{L}(z) > 1$.
}
the inverse right-handed neutrino decay $l_\alpha H \to N_1$ is in equilibrium.

We have numerically checked the range of $z$ where the inverse decay is in thermal equilibrium. This range can be expressed as
\begin{align}
\label{eq: SM equilibrium range}
z_{\mr{in}}^{L} < z < z_{\mr{fo}}^{L},
\end{align}
where $z_{\mr{in}}^{L} \simeq 0.5$ and $z_{\mr{fo}}^{L} \simeq 10$, which is consistent with the previous work~\cite{Chun:2023eqc}.

Following Refs.~\cite{Co:2020xlh, Chun:2023eqc}, we regard
\begin{align}
\label{eq: conserved charge}
Y_\theta := f^2 \dot{\theta}(T)/s(T),
\end{align}
where $s(T)$ is the entropy density, as a conserved parameter. 
Then, the baryon (lepton) asymmetry $Y_B := n_{\Delta B}/s$ ($Y_L := n_{\Delta L}/s$) is determined by the value of $\dot{\theta}$ at the time when the inverse decay process decouples, and remains conserved thereafter. 
Thus, the resulting baryon (lepton) asymmetry is given by
\begin{align}
Y_{B}(z_{\rm fo}^L) 
    &= \frac{c_B}{6} \, Y_{\theta} \, \frac{g_{N,1}^2}{(\sqrt{2} \, z_{\mathrm{fo}}^L)^2}, \\
Y_{L}(z_{\rm fo}^L) 
    &= \frac{c_L}{6} \, Y_{\theta} \, \frac{g_{N,1}^2}{(\sqrt{2} \, z_{\mathrm{fo}}^L)^2}.
\end{align}

\begin{figure}[t]
  \centering
  \includegraphics[width=0.45\textwidth]{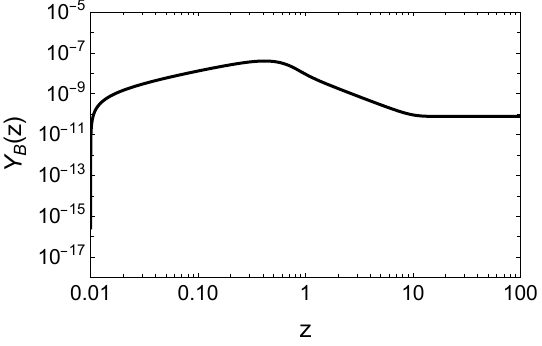}
  \caption{Evolution of the baryon asymmetry generated through inverse decay in the presence of a majoron background. 
  The mass and Yukawa coupling of the right-handed neutrino are fixed at 
  $(M_1, y_{N,1}) = (3 \times 10^5 ~{\rm GeV}, 2 \times 10^{-5})$, which yields $K \simeq 50$.
  To reproduce the observed baryon asymmetry, we set  $Y_\theta \simeq 10^{-7} g_{N,1}^{-2} (z_{\mr{fo}}^L/10)^2 $.}
  \label{fig: Solution for BAU Boltzmann equation}
\end{figure}

To reproduce the observed baryon asymmetry, $Y_{B,\,\mathrm{obs}} \simeq 8.7 \times 10^{-11}$, we require $Y_\theta \simeq 10^{-7} g_{N,1}^{-2} (z_{\mr{fo}}^L/10)^2 $. 
In Fig.~\ref{fig: Solution for BAU Boltzmann equation}, we show the numerical solution for the produced baryon asymmetry, which follows from Eq.~\eqref{eq: Boltzmann equation for lepton sector}. 

We note that, since the electroweak sphaleron process becomes decoupled when $T \lesssim 130~\mathrm{GeV}$~\cite{DOnofrio:2014rug}, the BAU is fixed by the sphaleron decoupling temperature if $M_1 / z_{\mathrm{fo}}^L \lesssim 130~\mathrm{GeV}$.

\section{Dark matter production}
\label{sec: dark matter production}

In this section, we discuss the spontaneous cogenesis of our scenario, focusing on DM production. 

In our model, the dark-sector particles $\chi$ and $\phi$ can also undergo inverse decay into right-handed neutrinos through the Yukawa interaction with coupling $|y_{\chi,1}|$. This inverse decay yields a DM asymmetry analogous to the SM lepton asymmetry. The Boltzmann equation for DM asymmetry $n_{\Delta \chi}:= n_{\chi}-n_{\bar{\chi}}$ in a $\dot{\theta}$ background equations are given by
\begin{align}
\label{eq: Boltzmann equation for dark sector}
\dot{n}_{\Delta \chi} + 3 H n_{\Delta \chi} = - n_{N_1}^{\rm{eq}}\langle{\Gamma_{N_1 \to \chi \phi} \rangle} \left(\frac{n_{\Delta \chi }}{n_{\chi}^{\rm{eq}}} +\frac{n_{\Delta \phi}}{n_\phi^{\rm{eq}}} -\frac{\dot{\theta}}{T}\right),
\end{align}
where $n_{\Delta \phi} := n_{\phi} - n_{\phi^*}$, $n_{X}^{\rm eq}, X=\chi, \phi$ represent the equilibrium number density of $X$, and
\begin{align}
\langle{\Gamma_{N_1 \to \chi \phi} \rangle}:= {K_1(z)\over K_2(z)}\Gamma_{N_1 \to \chi \phi}.
\end{align}
Again, in Eq~\eqref{eq: Boltzmann equation for dark sector}, we neglect the effect of the CP-violating decay process.

Besides decays and inverse decays, scattering processes can also produce or wash out dark matter asymmetry under a kinetic majoron background. These include $\Delta L = 2$ scattering processes that conserve the SM lepton number, such as $\chi\phi \to \bar{\chi}\phi^{*}$, and scattering processes that violate the SM lepton number, such as $\chi\phi \leftrightarrow L H$. 
In addition, we have $\Delta L = 1$ processes such that $ \chi \phi \to N_1 Z'$ and $ \chi Z' \to N_1 \phi$, but these processes are subleading effects compared to the inverse decay. 
Among these, the first processes can be important, particularly in the low-temperature regime $z \gg 1$, unless $y_{\chi,1}$ is sufficiently small.\footnote{
From the seesaw relation, once the right-handed neutrino mass scale is fixed, the typical value of $y_{N,1}$ can be estimated as
\begin{equation}
y_{N,1}^2 \simeq \frac{m_\nu M_1}{v^2},
\end{equation}
where $m_\nu \simeq 0.05~\mr{eV}$ is the SM neutrino mass scale and $v = 174~\mr{GeV}$ is the Higgs VEV. If we focus on low-scale leptogenesis with $M_1 \ll 10^{14}~\mr{GeV}$, the coupling $y_{N,1}$ is therefore quite small.

On the other hand, $y_{\chi,1}$ is not constrained by the seesaw relation, and thus it can take relatively large values even when the right-handed neutrino mass scale is small.}

The reason is as follows. In the temperature regime where the right-handed neutrino becomes non-relativistic, $z\gg 1$, inverse decays become almost negligible due to the Boltzmann suppression. By contrast, the $\Delta L = 2$ scattering processes with an intermediate right-handed neutrino are merely scatterings suppressed by the cutoff scale $M_1$. Their interaction rates, therefore, scale as a power law in temperature. As a result, if the coupling $y_{\chi,1}$ is sufficiently large, these scatterings can remain in equilibrium even after the inverse decays have decoupled. The latter scattering processes that violate the SM lepton number may also become important in the low-temperature regime $z \gg 1$ for the same reason. However, compared with scatterings within the dark sector, they are suppressed by a factor of $|y_{N,\alpha1}|^2$. Therefore, a larger coupling $y_{\chi,1}$ is required for their effects to become significant.

In this work, however, we restrict ourselves to the regime of small couplings, $|y_{\chi,1}| < y_{\chi,1}^{\mathrm{sct}}$, where scattering can safely be ignored. 
Here, $y_{\chi,1}^{\mathrm{sct}}$ denotes the critical coupling defined by the condition that the inverse decays go out of equilibrium at the same value of $z$ at which the $\Delta L = 2$ scatterings, which can produce additional dark matter asymmetry, decouple.
From a numerical comparison of the Hubble parameter with the interaction rates of $\Delta L = 2$ scatterings and inverse decay processes within the dark sector, we find
\begin{align}
  M_1 = 3 \times 10^{3}\,\mathrm{GeV} &:\quad y_{\chi,1}^{\mathrm{sct}} \simeq 2 \times 10^{-3}, \\
  M_1 = 3 \times 10^{5}\,\mathrm{GeV} &:\quad y_{\chi,1}^{\mathrm{sct}} \simeq 7 \times 10^{-3} .
\end{align}

\begin{figure}
  \centering
  \subcaptionbox{\label{fig: RateMN3TeV}
   $M_1 = 300$ TeV, $y_{N,1} = 2\times 10^{-5}$, and $|y_{\chi,1}| =10^{-5}$.
  }
  {\includegraphics[height=50mm]{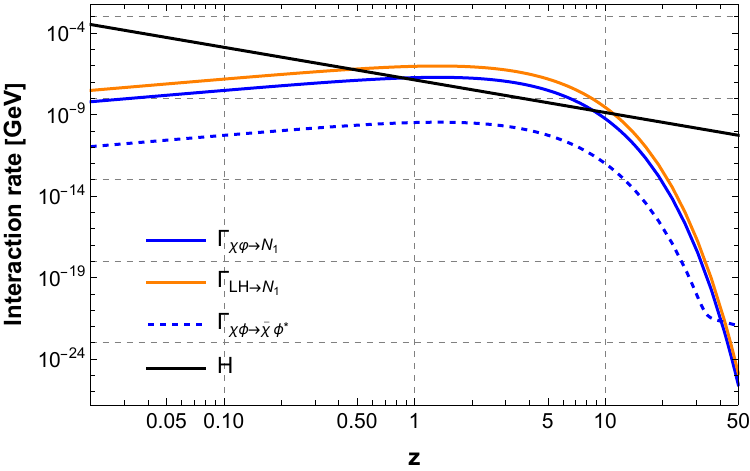}}\\\vspace{.5cm}
  \subcaptionbox{\label{fig: RateMN300TeV}
  $M_1 = 300$ TeV, $y_{N,1} = 2\times 10^{-5}$, and $|y_{\chi,1}| = y_{\chi,1}^{\rm sct} \simeq 7\times 10^{-3}$.
  }
  { 
  \includegraphics[height=50mm]{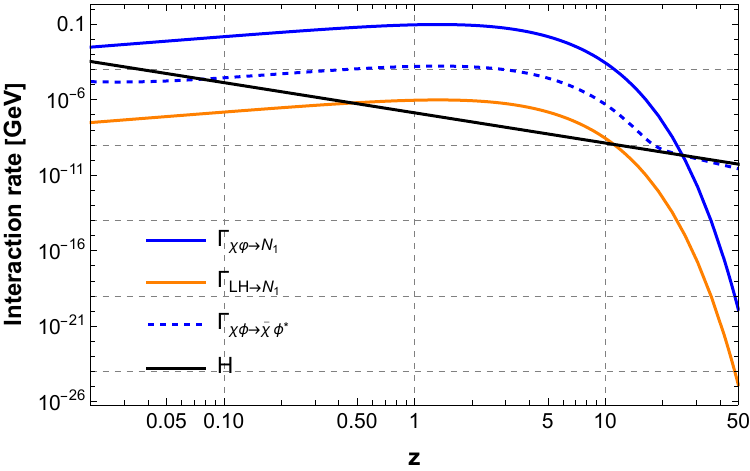}}
  \caption{The interaction rates of $L H \to N_1$, $\chi\phi \to N_1$, and $\chi\phi\to\bar\chi\phi^*$, and the Hubble rate are shown as functions of $z$.
  \label{fig: interactionrate}}
  \end{figure}

To understand the production of the asymmetry, it is important to understand when (and in what order) the interactions responsible for producing the asymmetry decouple.
To see this, in Fig.~\ref{fig: interactionrate} we show the interaction rates of the inverse decays in the SM and in the dark sector, as well as that of $\chi \phi \to \bar{\chi} \phi^{*}$, which serves as a representative process of the $\Delta L = 2$ scatterings, which can produce dark matter asymmetry.\footnote{
For lepton number-violating scatterings, the scattering rate increases with $z$ for $z \lesssim 1$. Indeed, since lepton-number violation requires inserting the Majorana mass, one can estimate
\begin{align}
\Gamma(\chi \phi \to \bar{\chi}\,\phi^*)
    \sim |y_{\chi,1}|^4\, \frac{M_1^{\,2}}{T} \propto z \, ,
\end{align}
but we note that obtaining the correct quantitative behavior requires subtracting the on-shell resonance contribution, as stated in the main text.
}

In Fig.~\ref{fig: interactionrate}, we fix $M_1 = 3 \times 10^5~\mr{GeV}$ and $y_{N,1} = 2 \times 10^{-5}$ (which corresponds to $K \simeq 50$) as representative values, and compare the behavior of the interaction rates as the value of $|y_{\chi,1}|$ is varied. In the upper panel, we take $|y_{\chi,1}| = 10^{-5}$ as the benchmark, while in the lower panel we also adopt $|y_{\chi,1}| = y_{\chi,1}^{\rm sct} \simeq 7\times 10^{-3}$. In addition, for simplicity, we treat all particles except the right-handed neutrino as massless.

As is evident from these figures, the magnitude of $|y_{\chi,1}|$ not only changes the interaction rate of the inverse decay $\chi \phi \to N_1$ into the dark sector, but also shifts the timing at which the process decouples. We note that for $z \lesssim 1$, the scattering process is dominated by the $s$-channel contribution $\chi \phi \to \bar{\chi}\,\phi^*$ that contains an on-shell resonance. In plotting Fig.~\ref{fig: interactionrate}, we show the scattering rate after performing the appropriate subtraction to avoid double counting with the decay contribution. Further details are provided in Appendix~\ref{app: delta 2 scattering rate}. 
For $z \gg 1$, the particles in the thermal bath can no longer hit the on-shell resonance of the right-handed neutrino, which means that scattering processes effectively correspond to those obtained by an effective interaction after integrating out the right-handed neutrino. Thus, the interaction rate of the $\Delta L = 2$ scatterings exhibits a transition to a linear behavior beyond a certain point of $z$.

As a result of this behavior, if $|y_{\chi,1}|$ is taken to be sufficiently larger than $y_{\chi,1}^{\rm sct}$, the $\Delta L = 2$ scatterings remain in thermal equilibrium for a relatively extended period, until the temperature drops below the masses of the dark-sector particles, $m_{\chi}$ and $m_{\phi}$, as we discussed before. During this stage, the dark matter asymmetry is produced in the presence of the Majoron background, and the resulting final asymmetry can no longer be regarded as being dominated by that generated through inverse decays.

To compute the produced DM asymmetry, it is important to determine when the inverse decay from dark sector particles is in thermal equilibrium. For this purpose, we define the dark sector version of Eq.~\eqref{eq: inverse decay criterion for leptogenesis}:
\begin{align}
\label{eq: inverse decay criterion for darkgenesis}
W_{\rm{ID}}^{D}(z)
:=z \frac{\langle{\Gamma_{N_1 \to \chi \phi}\rangle}}{H_1}\frac{n^{\rm{eq}}_{N_1}}{n^{\rm{eq}}_{\chi}}.
\end{align}
If $z W_{\rm{ID}}^{D}(z) > 1$, inverse decay of right-handed neutrino, $\chi \phi \to N_1$ is in equilibrium.
The range of $z$ where $z W_{\rm{ID}}^{D}(z) > 1$ is approximately expressed as
\begin{align}
\label{eq: DM equilibrium range}
z_{\mr{in}}^{D} < z < z_{\mr{fo}}^{D},
\end{align}
where
\begin{align}
&z_{\mr{in}}^{D}  
    \simeq 0.7 \left(M_1 \over 10^5 ~\mr{GeV}\right)\left(10^{-5} \over |y_{\chi,1}|\right)^2,\\
\label{eq: zfoD solution}
&z_{\mr{fo}}^{D}  
      \simeq  {- 7 \over 2} W_{-1} \Bigg( {- 2 \over 7} \exp \Bigg[{20 \over 7} \sqrt{\pi  \over 2} \nonumber  \left(|y_{\chi,1}| \over 10^{-5} \right)^2  \left( 10^5~\mr{GeV} \over M_1 \right)\Bigg]\Bigg), \\
\end{align}
with $W_{-1}(x)$ denoting product logarithm function (Lambert W function). For example, we obtain $z_{fo}^{D} \simeq 10$, if we take $(M_1, |y_{\chi,1}|) = (10^5~\mr{GeV}, 10^{-5})$.

As mentioned above, the Yukawa coupling $y_{\chi,1}$ can be freely chosen. Therefore, we note that inverse decay from dark sector particles is not necessarily in equilibrium in thermal history. This allows us to consider two types of scenarios for DM asymmetry production: the freeze-out and the freeze-in scenarios.

\begin{figure}[t]
  \centering
  \includegraphics[width=0.5\textwidth]{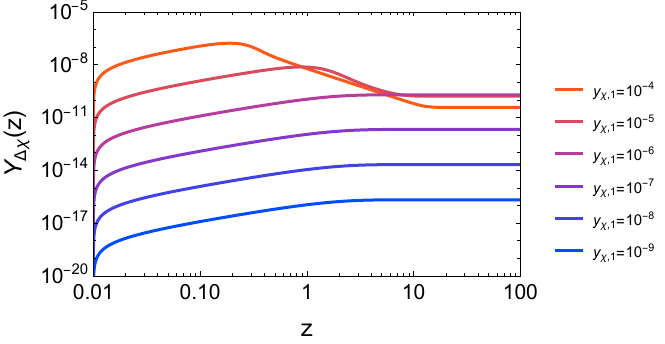}
  \caption{Evolution of the dark matter asymmetry generated through inverse decay in the presence of a majoron background, shown for different values of $|y_{\chi,1}|$. 
The mass and Yukawa coupling of the right-handed neutrino, as well as the value of $Y_\theta$, are taken to be the same as in Fig.~\ref{fig: Solution for BAU Boltzmann equation}. 
The final value of $Y_{\Delta \chi}$ reaches its maximum when $|y_{\chi,1}|$ lies at the boundary between the freeze-out and freeze-in regimes of the inverse decay. 
In the freeze-out regime, larger $|y_{\chi,1}|$ leads to a smaller $Y_{\Delta \chi}$, while in the freeze-in regime, smaller $|y_{\chi,1}|$ results in a smaller $Y_{\Delta \chi}$.
}
  \label{fig: Solution for DM Boltzmann equation}
\end{figure}

\subsection{Freeze-out}
First, let us consider the situation where the inverse decay of dark-sector particles is in equilibrium, that is, the regime in which $z_\mathrm{in}^D < z_\mathrm{fo}^D$ holds.  
In this case, the asymmetric component of the DM number density is given by
\begin{align}
\label{eq: DM asymmetry in terms of number density}
    n_{\Delta \chi} = \frac{c_{\chi}}{6} \dot{\theta} T^2,
\end{align}
as long as the inverse decay is in thermal equilibrium. Here, $c_{\chi}$ is the coefficient determined by the chemical equilibrium conditions.  
Assuming that $\phi$ carries no asymmetry due to the presence of sufficiently rapid interactions that interchange $\phi$ and $ \phi^*$, the coefficient is given by $c_\chi \simeq 1/2$.
\footnote{
For example, the $\phi$ asymmetry can be washed out if there exists 
the following interaction:
\begin{align}
\mathcal{L} \supset g_{\Psi}\,\phi\,\bar{\Psi}\Psi + h.c.,
\end{align}
where $\Psi$ carries a conserved charge $Q_{\Psi}$. If this interaction is in chemical equilibrium, the chemical potentials of both $\Psi$ and $\phi$ vanish. Another realization is a mass-mixing term, 
\begin{align}
\mathcal{L} \supset \mu^2 (\phi \phi+\phi^* \phi^*),
\end{align}
which induces $\phi-\phi^*$ oscillations and washes out the $\phi$ asymmetry~\cite{Buckley:2011ye, Tulin:2012re, Nagata:2016knk, Ghosh:2021qbo}. Hence, if either of these effects is in thermal equilibrium during leptogenesis and proceeds sufficiently faster than $\chi\phi \to N_1$, the asymmetry stored in $\phi$ can be neglected.}

As we can see in Eq.~\eqref{eq: DM asymmetry in terms of number density}, the dark-matter asymmetry in equilibrium coincides with that of the baryon or lepton asymmetry (See Eqs~\eqref{eq: baryon asymmetry in terms of number density} and \eqref{eq: lepton asymmetry in terms of number density}), except for the coefficients determined by the equilibrium conditions. This is because the CP violation does not originate from the CP phase parameters in the model, but rather from the single majoron background $J =f\theta $.

In terms of the dark-matter asymmetry $Y_{\Delta \chi}(z) := n_{\Delta \chi}/s$,  the resulting DM abundance is given by
\begin{align}
\label{eq: DM asymmetry}
Y_{\Delta \chi}(z_{\rm fo}^D)
    &= \frac{c_\chi}{6}\, Y_{\theta}\, \frac{g_{N,1}^2}{(\sqrt{2}\, z_{\mathrm{fo}}^D)^2},
\end{align}
which follows from Eq.~\eqref{eq: DM asymmetry in terms of number density}. $Y_\theta$ is defined in Eq.~\eqref{eq: conserved charge}.

In Fig.~\ref{fig: Solution for DM Boltzmann equation}, we show the numerical solution of the $Y_{\Delta \chi}$, obtained by solving Eq.~\eqref{eq: Boltzmann equation for dark sector}, for different choices of $|y_{\chi,1}|$. 
In the figure, as one of the benchmark points, we adopt $(M_1, y_{N,1}) = (3 \times 10^5~ \rm{GeV}, 2 \times 10^{-5})$, the same benchmark as in Fig.~\ref{fig: Solution for BAU Boltzmann equation}.

In the freeze-out regime, where $Y_{\Delta \chi}$ tracks its equilibrium value, we find that taking a larger $|y_{\chi,1}|$ results in a smaller final asymmetry. 
This is because a larger $|y_{\chi,1}|$ delays the departure from equilibrium, so that the dynamical background CP phase has already experienced redshift, $\dot{\theta} \propto T^3$, at the time of decoupling.
Indeed, as $|y_{\chi,1}|$ increases, the freeze-out parameter $z_{\mathrm{fo}}^D$ becomes larger, and consequently the generated asymmetry is suppressed according to Eq.~\eqref{eq: DM asymmetry}.

In the freeze-out scenario, using the analytic expression~\eqref{eq: DM asymmetry}, the DM mass is given by
\begin{align}
m_\chi \simeq \frac{\Omega_{\mathrm{DM}}}{\Omega_B} 
\left( \frac{c_B}{c_\chi} \right) 
\left( \frac{z_{\mathrm{fo}}^D}{z_{\mathrm{fo}}^L} \right)^2 m_p .
\end{align}
Here, $\Omega_{\mathrm{DM}} = \rho_{\mathrm{DM}} / \rho_{\mathrm{crit}}$ and 
$\Omega_B = \rho_B / \rho_{\mathrm{crit}}$ denotes the present DM and baryon energy densities, normalized by the critical density $\rho_{\mathrm{crit}}$ respectively. 
The ratio of the energy densities of DM to baryons is approximately 
${\Omega_{\mathrm{DM}}}/{\Omega_B} \simeq 5.4$~\cite{Planck:2018vyg}, and $m_p$ is the proton mass.

\begin{figure}[t]
  \centering
  \includegraphics[width=0.45\textwidth]{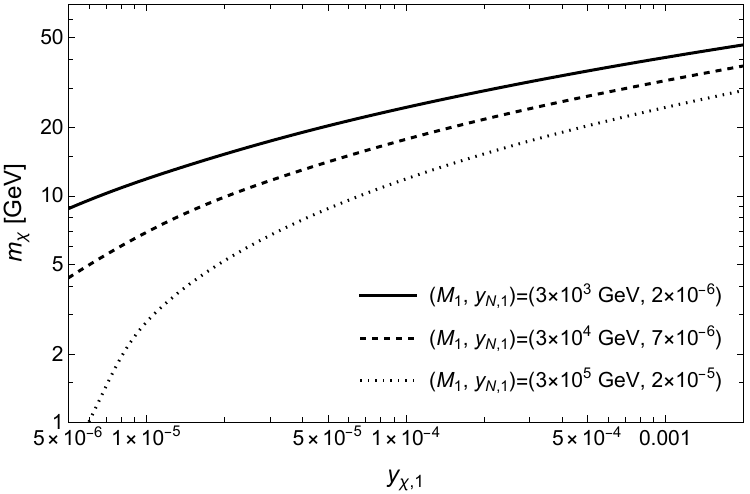}
  \caption{The dark matter mass $m_{\chi}$ required to account for the observed relic abundance in the freeze-out scenario. As benchmark points, we choose $(M_1, y_{N,1}) = (3 \times 10^3~{\rm GeV}, 2.2 \times 10^{-6})$ (black solid line), $(3 \times 10^4~{\rm GeV}, 7 \times 10^{-6})$ (dashed line), and $(3 \times 10^5~{\rm GeV}, 2 \times 10^{-5})$ (dotted line).
}
  \label{fig: Dark matter mass in freeze out}
\end{figure}

Figure~\ref{fig: Dark matter mass in freeze out} shows the relation between the DM mass and $|y_{\chi,1}|$, obtained by numerically solving the Boltzmann equations. 
As benchmarks, we selected three pairs of the right-handed neutrino mass and the corresponding $y_{N,1}$ from the strong wash-out condition $K \simeq 50$, which are also indicated in the figure. 
The $|y_{\chi,1}|$ dependence of $z_{\rm fo}^{D}$ given in Eq.~\eqref{eq: zfoD solution} can be well approximated by $\log |y_{\chi,1}|^2$, and the DM mass required to reproduce the observed relic abundance increases approximately in proportion to its square.

Conversely, as $|y_{\chi,1}|$ decreases, the DM mass that reproduces the observed relic abundance also decreases. Eventually, this reaches a minimum value at the boundary between the freeze-out and freeze-in regimes. 
We numerically confirm that the minimum value of $m_{\chi}$ is 
\begin{align}
m_{\chi, \mathrm{min}} \simeq 0.5~\mathrm{GeV}.
\end{align}
For even smaller $|y_{\chi,1}|$, the system enters the freeze-in regime, where, as will be discussed later, the DM mass required to account for the correct relic abundance starts to increase as $|y_{\chi,1}|$ becomes smaller.

\subsection{Freeze-in}
Next, we consider the case where the inverse decay from dark sector particles is never in thermal equilibrium, while the dark sector particles are thermalized with SM particles.\footnote{
In general, the temperature for the dark sector particles $T_d$ is lower than that of the SM, $T$; namely, $T_d := \xi T$ with $\xi \leq 1$. Since the temperature ratio between the two sectors depends on the model, we analyze the case with $\xi = 1$ in what follows.
}
Unlike the freeze-out scenario, the produced DM asymmetry is smaller than that of the ``equilibrium'' value $n_{\Delta \chi} \ll \dot{\theta} T^2$ in the freeze-in scenario. Therefore, Eq.~\eqref{eq: Boltzmann equation for dark sector} can be reduced to
\begin{align}
\label{eq: Boltzmann equation for freeze in dark sector}
\dot{n}_{\Delta \chi} + 3 H n_{\Delta \chi} \simeq n_{N}^{\rm eq} \langle \Gamma_{N \to \chi \phi} \rangle \frac{\dot{\theta}}{T},
\end{align}
in the freeze-in case. From this equation, we can see that when the asymmetry production terminates (i.e., $\dot{n}_{\Delta \chi} \simeq 0$), the number density at that time is suppressed by a factor of $W_{\rm ID}^{D}(z(T))$ compared to its “equilibrium” value, 
$n_{\Delta \chi}^{\rm eq}(T) := (c_{\chi}/6)\, \dot{\theta}(T) T^2$:
\begin{align}
n_{\Delta \chi}(T_{\rm sat}) 
    \simeq \left(M_1 \over T_{\rm sat}\right) W_{\rm ID}^{D}(z(T_{\rm sat})) \, n_{\Delta \chi}^{\rm eq}(T_{\rm sat}),
\end{align}
where $T_{\rm sat}$ is temperature when DM asymmetry production terminates. $W_{\rm ID}^{D}(z)$ is defined in Eq.~\eqref{eq: inverse decay criterion for darkgenesis}.

This suppression can also be obtained by solving Eq.~\eqref{eq: Boltzmann equation for freeze in dark sector} directly in terms of asymmetry $Y_{\Delta \chi}$:
\begin{align}
Y_{\Delta \chi}(z_{\rm sat})  
    &\simeq 
        \int^{z_{\rm sat}}_0 dz W_{\rm ID}^{D}(z) \, Y_{\Delta \chi}^{\rm eq}(z) \\   
    &\simeq 
        z_{\rm sat}W_{\rm ID}^{D}(z_{\rm sat}) \,  Y_{\Delta \chi}^{\rm eq}(z_{\rm sat}),
\end{align}
where $z_{\rm sat} := M_1/ T_{\rm sat}$ and  $Y_{\Delta \chi}^{\rm eq}(z) := c_\chi Y_{\theta} g_{N,1}^2 / (12 (z_{\mathrm{fo}}^D)^2)$, which is given by the right-hand side in Eq.~\eqref{eq: DM asymmetry}.
We note that the integrand becomes approximately constant for $z \lesssim 1$. Therefore, unless $z_{\rm sat} \gg 1$, it is a good approximation to evaluate it at $z \simeq z_{\rm sat}$ and take it out of the integral, and we have checked $z_{\rm sat} \simeq 1$ numerically.

Once $z$ exceeds $z_{\rm sat}$, the asymmetry saturates, \Ht{and} the present abundance is determined by $Y_{\Delta \chi}(z_{\rm sat})$. In Fig.~\ref{fig: Solution for DM Boltzmann equation}, we can see this saturated behavior for the numerical solution with small $|y_{\chi,1}|$.

In the freeze-in scenario, the DM mass is given by
\begin{align}
\label{eq: dark matter mass freeze in}
m_\chi \simeq \frac{\Omega_{\mathrm{DM}}}{\Omega_B} 
\left( \frac{c_B}{c_\chi} \right) 
\left( \frac{z_{\rm sat}}{z_{\mathrm{fo}}^L} \right)^2 {m_p \over z_{\rm sat}W_{\rm ID}^{D}(z_{\rm sat})} .
\end{align}
In contrast to the freeze-out scenario, the DM mass given in Eq.~\eqref{eq: dark matter mass freeze in} increases in proportion to the inverse square of $|y_{\chi,1}|$ as $|y_{\chi,1}|$ decreases. 
This behavior arises because the asymmetry has a suppression factor of $W_{\rm ID}^{D}(z) \propto |y_{\chi,1}|^2$. 

\begin{figure}[t]
  \centering
  \includegraphics[width=0.5\textwidth]{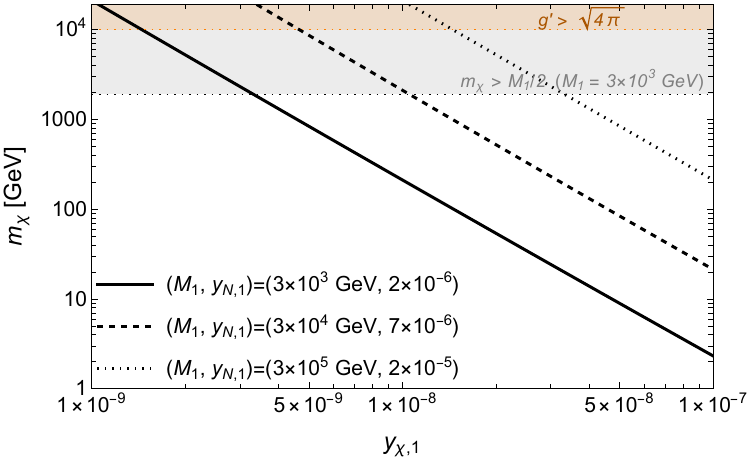}
  \caption{The required dark matter mass $m_\chi$ in the freeze-in scenario. The benchmark values are the same as those in Fig.~\ref{fig: Dark matter mass in freeze out}. In drawing the figure, we have neglected the dependence on $m_{\chi}$ and $m_{\phi}$; instead, the region where the required dark matter mass exceeds one-half of the right-handed neutrino mass is shaded. We also shade the region where the perturbativity condition $g^\prime < \sqrt{4\pi}$ is not satisfied.
}
  \label{fig: Dark matter mass in freeze in}
\end{figure}

In Fig.~\ref{fig: Dark matter mass in freeze in}, we show the relation between the DM mass $m_{\chi}$ and the Yukawa coupling $y_{\chi,1}$ for the same three benchmark choices as in Fig.~\ref{fig: Dark matter mass in freeze out}. 
To plot this figure, the dependence on $m_{\chi}$ and $m_{\phi}$ in the interaction rate of the inverse decay $\chi \phi \to N_1$ is neglected. 
In the figure, we shade the region where the perturbativity condition $g^\prime < \sqrt{4\pi}$ is not satisfied; as we will discuss in Sec.~\ref{sec: discussion}, the larger the dark matter mass becomes, the larger the size of $g^\prime$ becomes to remove the symmetric components of $\phi$ and $\chi$. In particular, for $m_{\phi} \gtrsim 10^{4}~\mr{GeV}$, the symmetric component of $\phi$ is no longer negligible compared to the asymmetric component of $\chi$.
Taking into account the assumption $m_{\chi} < m_{\phi}$, we shade in orange the corresponding region with $m_{\chi} \gtrsim 10^{4}~\mr{GeV}$.

We also shade the region where the required dark matter mass exceeds one-half of the right-handed neutrino mass, for $M_1 = 3\times 10^3$ GeV; in this region, the inverse decay is kinematically forbidden, and thus the asymmetric component of DM cannot be generated.
We note that, since $m_{\phi} > m_{\chi}$ in our scenario, this constraint is conservative, in the sense that for larger values of $m_{\phi}$, the constraints from this kinematics become more stringent.

Combining these constraints, for a given $M_{1}$, the value of $|y_{\chi,1}|$ at which the required DM mass intersects the shaded region represents the lower bound on $|y_{\chi,1}|$ that can realize ADM. 

In the freeze-in scenario, unlike the freeze-out case, the required value of $m_{\chi}$ is sensitive to $|y_{\chi,1}|$, and the ADM mechanism can be realized over a wider range of masses. In particular, when the mass of the right-handed neutrino is sufficiently large, the required value of $m_{\chi}$ can exceed the $\rm{TeV}$ scale, which is far beyond the DM mass scale for the case of the freeze-out.

\section{Discussions on phenomenology}
\label{sec: discussion}

\subsection{Majoron constraints}
In our scenario, the majoron plays an essential role as a background field. As already mentioned, successful low-scale leptogenesis requires the majoron field to undergo kinetic misalignment. On the other hand, if its kinetic energy becomes too large, it would modify the thermal history of the universe. Therefore, our analysis is valid only when  
$f^2 \dot{\theta}^2 (T_{\mr{fo}})/2 < \pi^2/30\, g_{*}(T_{\mr{fo}}^L) (T_{\mr{fo}}^L)^4$ holds,  where $T_{\mr{fo}}^L := M_1/ z_{\mr{fo}}^L$ and $g_{*}$ is the effective degrees of freedom. 
To achieve successful spontaneous leptogenesis, this condition gives a constraint
\begin{align}
g_{N,1} \gtrsim 10^{-7}
\left( \frac{g_{*}(T_{\mr{fo}}^L)}{100} \right)
\left( \frac{z_{\mr{fo}}^L}{10} \right),
\end{align}
which has already been discussed in Ref.~\cite{Chun:2023eqc}
.

In addition, when the lifetime of the majoron field is sufficiently long,  the observed abundance of DM would be altered due to the majoron energy density, ${\rho_{J} / s}$. The origin of ${\rho_{J} / s}$ can be attributed to two sources:  one arising from the kinetic misalignment mechanism~\cite{Co:2019jts, Chang:2019tvx} and the other from thermal production.

In the kinetic misalignment mechanism, the energy density of the majoron depends on the relation between  
the temperature $T_{\mr{osc}}$, at which the majoron mass becomes comparable to the Hubble parameter,  
and the temperature $T_{\mr{trap}}$, at which its kinetic energy equals the potential barrier.  
In each case, the energy density of the majoron is expressed as~\cite{Co:2019jts, Chang:2019tvx}
\begin{align}
{\rho_{J} \over s}
	\simeq 
\begin{cases}
{m_J^2 f^2 / s(T_{\mr{trap}})} \simeq m_J Y_{\theta} & (T_{\mr{osc}} > T_{\mr{trap}}), \\[4pt]
{m_J^2 f^2 / s(T_{\mr{osc}})} & (T_{\mr{osc}} < T_{\mr{trap}}),
\end{cases}
\end{align}
where $m_J$ denotes the majoron mass.

Except for the case where the majoron is extremely light,  
the condition $T_{\mr{osc}} > T_{\mr{trap}}$ is satisfied in our case,  
and hence the energy density is determined by the product of $m_J$ and $Y_{\theta}$.  
To account for the BAU, one needs  
$Y_\theta \simeq 10^{-7} g_{N,1}^{-2} (z_{\mr{fo}}^L/10)^2 $,  so that, for the majoron energy density to be negligible, the majoron mass must satisfy
\begin{align}
m_J \lesssim 10~\mr{meV} 
\left( \frac{g_{N,1}}{10^{-4}} \right)^2.
\end{align}
Next, we consider the thermal production of the majoron.  
When the following interactions,
\begin{align}
LH &\leftrightarrow N_i J, \\
\chi \phi &\leftrightarrow N_i J,
\end{align}
maintain equilibrium at $T > M_1$, these interaction leads to the thermal production of the majoron.  

In the freeze-out scenario, where $|y_{\chi,1}| > y_{N,1}$, the latter process $\chi \phi \leftrightarrow N_i J$ dominates.  
The requirement that the energy density of the majoron remains small imposes the following constraint on $g_{N,1}$:
\begin{align}
g_{N,1}
	\lesssim
	10^{-4}
	\left( \frac{M_1}{10^{6}~\mr{GeV}} \right)^{1/2}
	\left( \frac{10^{-3}}{|y_{\chi,1}|} \right).
\end{align}
On the other hand, in the freeze-in scenario, where $|y_{\chi,1}| < y_{N,1}$,  
the former process $LH \leftrightarrow N_i J$ becomes dominant,  
which requires $g_{N,1} \lesssim 0.1$~\cite{Chun:2023eqc}.

\subsection{ADM constraints}

In ADM scenarios, the reduction of the symmetric component is one of the key issues. As the most minimal annihilation channel, one may consider the process $\chi \bar{\chi} \to \phi \phi^{*}$ induced by the Yukawa interaction essential for generating the asymmetry, $\mathcal{L} \supset y_{\chi,1} \bar\chi \phi^{\dagger} N_1$. However, as shown in Appendix~\ref{appendix: symmetric component}, this process is almost irrelevant for removing the symmetric component.
As another possibility, one may consider the annihilation channel $\chi \bar{\chi} \to n \bar{n}$ mediated by the nucleon interaction 
\begin{align}
\label{eq: effective interaction with nucleous}
\mathcal{L}_{\text{int}} \supset \frac{1}{\Lambda^2}\, \chi \bar{\chi}\, n \bar{n},
\end{align}
where $n$ denotes a nucleon and $\Lambda$ is a cut-off scale. However, such a possibility is excluded by constraints from direct detection experiments and collider searches in the range $1~\mr{GeV} \lesssim m_{\chi} \lesssim 100~\mr{GeV}$~\cite{Buckley:2011kk, March-Russell:2012elz}, which coincides with the typical mass range predicted by our freeze-out scenario.

\begin{figure*}[t]
  \centering
  \includegraphics[height=50mm]{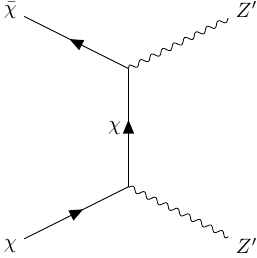}
  \hspace{1cm}
  \includegraphics[height=50mm]{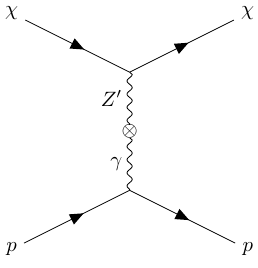}
  \caption{Feynman diagrams for the dark matter annihilation (left) and direct detection signals (right), respectively.}
  \label{fig: diagrams}
\end{figure*}

As the final possibility, the DM may annihilate into dark-sector particles through another interaction. In our Lagrangian~\eqref{Eq: Lagrangian}, a dark photon is introduced, which leads to the annihilation process $\chi \bar{\chi} \to Z' Z'$ shown in Fig.~\ref{fig: diagrams}.\footnote{
The process $\chi \bar{\chi} \to Z' Z'$, referred to as secluded annihilation, is a channel that opens only when $m_{\chi} > m_{Z'}$. Conversely, when $m_{\chi} < m_{Z'}$, annihilation proceeds through the direct annihilation channel $\chi \bar{\chi} \to f \bar{f}$ mediated by the $Z'$, where $f$ denotes an SM fermion. However, the latter effectively reduces to the operator in Eq.~\eqref{eq: effective interaction with nucleous}, and is therefore already excluded by experiments.
} 
Requiring this process to efficiently remove the symmetric component of DM imposes a lower bound on the dark gauge coupling,
\begin{align}\label{eq: depletion of sym chi}
g' \gtrsim 
10^{-2}
\left( \frac{m_{\chi}}{1~\mr{GeV}} \right)^{1/2}.
\end{align}
Given the perturbativity condition, $g^\prime < \sqrt{4\pi}$, this leads to an upper bound on $m_\chi$, given by
\begin{align}\label{eq: perturbativity}
    m_\chi \lesssim 10^5~\text{GeV}.
\end{align}

The lower bound on the dark gauge coupling $g'$ for depleting the symmetric components Eq.~\eqref{eq: depletion of sym chi} (and also in Eq.~\eqref{eq: depletion of sym phi} for the depletion of $\phi$), would also lead to an upper bound on the kinetic mixing $\epsilon$, combined with constraints from direct detection experiments. 
As shown in Fig.~\ref{fig: diagrams}, the direct detection cross section in our model depends on $\epsilon$ and $g'$:
\begin{align}
    &\sigma_{\chi N\to\chi N} \simeq \sigma_{\chi p\to\chi p} \frac{\mu_{\chi N}^2}{\mu_{\chi p}^2} \left(\frac{Z}{A}\right)^2,\\
    &\sigma_{\chi p\to\chi p} \simeq \frac{\epsilon^2 e^2{g^\prime}^2\mu_{\chi p}^2}{\pi m_{Z'}^4},
\end{align}
where $\mu_{\chi p}$ and $\mu_{\chi N}$ are the reduced masses for $\chi$ and the proton $p$ and nucleon $N$, respectively; $Z$ and $A$ denote the atomic and mass numbers of the target nucleus.
Therefore, given a direct detection bound on $\sigma_{\chi N\to\chi N} \lesssim \sigma_{\chi N\to\chi N}^{\rm exp
}$ for a DM mass $m_\chi$, this, together with the bounds on the coupling, Eq.~\eqref{eq: depletion of sym chi}, leads to
\begin{align}
    \epsilon 
    &\lesssim 10^{-8} ~ \left(\frac{\sigma_{\chi N\to\chi N}^{\rm exp
}}{10^{-44}~{\rm cm}^2}\right)^{1/2} \left(\frac{m_{Z'}}{0.1 ~{\rm GeV}}\right)^2 \\ \nonumber 
&\qquad\qquad \times \left(\frac{m_\chi}{1~ 
{\rm GeV}}\right)^{-1/2}\left(\frac{\mu_{\chi N}}{1~ 
{\rm GeV}}\right)^{-1} \left(\frac{Z}{A}\right)^{-1}.
\end{align}
Such an analysis has been conducted in Ref.~\cite{delaVega:2023dmw} for the case of thermal relic DM, based on the nuclear recoil direct detection results from PandaX–4T~\cite{PandaX-4T:2021bab}, XENON–1T~\cite{XENON:2018voc}, and PICO–60~\cite{PICO:2015pux,PICO:2019vsc}.
It is shown that for $m_{Z'} \gtrsim 10\,\mathrm{MeV}$, the resulting bound on $\epsilon$ gives stronger constraints on the size of kinetic mixing than the direct searches.\footnote{We note that since ADM requires a larger size of the dark gauge coupling than the thermal freezeout, the resulting constraint is also expected to be stronger than that shown in Ref.~\cite{delaVega:2023dmw}. Moreover, incorporating the results from LZ~\cite{LZ:2024zvo} and CRESST~\cite{CRESST:2019jnq,CRESST:2024cpr} would also strengthen the constraints discussed in Ref~\cite{delaVega:2023dmw}.}

One can also see that the dark photon portal DM models are quite constrained, while the remaining region of the parameter space can be a prime target for next-generation experiments.

\subsection{Constraints on scalar $\phi$}
Another constraint on the model parameters is obtained by requiring that the energy density of $\phi$ is also negligibly small.
Since the asymmetry of $\phi$ is efficiently washed out in our scenario, it is the symmetric component that can be a dominant source of energy density, and its abundance is determined by the freezeout of the process $\phi \phi^* \to Z' Z'$. 
The ratio of the abundance of $\phi$ to that of the observed dark matter is given by 
\begin{align}
	\frac{\Omega_\phi}{\Omega_{\rm DM}} \sim 10^{-2} \left(\frac{g^\prime}{10^{-1}}\right)^{-4}  \left(\frac{m_\phi}{10 ~{\rm GeV}}\right)^{2},
\end{align}
where we assume $x_f \sim \mathcal{O}(10)$.
The condition for the process $\phi \phi^* \to Z' Z'$ to efficiently remove the symmetric component of $\phi$ gives
\begin{align}\label{eq: depletion of sym phi}
    g' \gg  
10^{-2}
\left( \frac{m_{\phi}}{10~\mr{GeV}} \right)^{1/2},
\end{align} 
which, given the perturbativity condition, leads to
\begin{align}
    m_\phi \lesssim 10^4~\text{GeV}.
\end{align}
Since we assume $m_\chi < m_\phi$, this also implies $m_\chi \lesssim 10^4$ GeV, which gives a stronger bound on $m_\chi$ than Eq.~\eqref{eq: perturbativity}.

In our model, $\phi$ can decay to $\bar\chi$ and $\bar\nu$, mediated by a right-handed neutrino.
Such decays to neutrinos are weakly constrained by Big Bang Nucleosynthesis (BBN), Cosmic Microwave Background (CMB), and diffuse neutrino/gamma fluxes~\cite{Kanzaki:2007pd}, depending on the lifetime. In our case, since the abundance of $\phi$ is guaranteed to be subdominant to the DM density, the lifetime of $\phi$, $\tau_\phi$,  can be as late as $\tau_\phi\lesssim 10^{12}$ s~\cite{Kanzaki:2007pd}.
Assuming that the masses of $\chi$ and $\phi$ are not so degenerate that the phase space factor is $\mathcal{O}(1)$, $\tau_\phi$ is given by\footnote{Despite the suppression from its phase space, the three-body decay $\phi \to \bar{\nu} \bar{\chi}h$ can in principle be dominant~\cite{Falkowski:2011xh}: the ratio of the decay rate, in the limit of $m_\phi \gg v$, is given by
\begin{align}
    \frac{\Gamma(\phi \to \bar{\nu} \bar{\chi}h)}{\Gamma(\phi \to \bar{\nu} \bar{\chi})} \simeq \frac{m_\phi^2}{24 \pi ^2v^2},
\end{align} 
which shows that for $m_\phi \gtrsim 10^4$ GeV, the three-body decay becomes dominant. Such a heavy mass, however, is not consistent with the perturbativity condition, and so the two-body decay is always dominant in our scenario.}
\begin{align}
    \tau_\phi 
    &\simeq 7 \times 10^{-3}~{\rm s}~\left(\frac{|y_{\chi,1}|}{10^{-3}}\right)^{-2} \left(\frac{m_\nu}{0.05~ {\rm eV}}\right)^{-1}\nonumber \\
    &\qquad \qquad \qquad\times \left(\frac{M_1}{10^5 ~{\rm GeV}}\right) \left(\frac{m_\phi}{10 ~{\rm GeV}}\right)^{-1},
\end{align}
and thus $\tau_\phi\lesssim 10^{12}$ is always satisfied in the parameter regions of interest.

\subsection{Dark photon constraints}
Next, we consider the constraints on the dark photon. These constraints differ depending on whether the dark photon can decay into an electron--positron pair.

We first consider the case in which the decay into an electron--positron pair is kinematically allowed. In this case, the dark photons produced from DM annihilation must decay before the epoch of BBN; otherwise, they would spoil the success of BBN or affect the CMB observations. To avoid this, the lifetime of the dark photon $Z'$ must be sufficiently short, such that it decays into an electron pair by the time when the temperature is in the $\mr{MeV}$ scale. This requirement imposes a lower bound on the kinetic mixing with the photon~\cite{Ruderman:2009tj},
\begin{align}
\epsilon \gtrsim 3 \times 10^{-11} \left( \frac{1~\mr{GeV}}{m_{Z'}} \right)^{1/2}.
\end{align}

In addition to the above constraint, for $m_{Z'}>2m_e$, collider experiments and B factories have already excluded $\epsilon e > 10^{-3}$ over a wide mass range, where $e$ denotes the electric charge~\cite{Bauer:2018onh}. 
For smaller kinetic mixing parameters, searches using fixed-target experiments are ongoing. 
For example, the NA64 collaboration has recently reported results from a search for dark photons~\cite{NA64:2023wbi}. 
For a more detailed overview of the current experimental searches and the projected future exclusion regions, see Ref.~\cite{Bauer:2018onh}.

We next consider the case $m_{Z'}<2m_e$, in which the decay into an electron--positron pair is kinematically forbidden.
If the dark photon is sufficiently light that the dark sector and the SM sector remain thermally coupled until $T \simeq \mathcal{O}(1)$ MeV, the effective number of relativistic degrees of freedom would affect the BBN and CMB observables. To avoid this, the interaction $Z' e^{\pm} \leftrightarrow \gamma e^{\pm}$ must have already decoupled by $T \simeq \mathcal{O}(1)$ MeV. This requirement leads to the constraint~\cite{Falkowski:2011xh},
\begin{align}
\epsilon \lesssim 
7 \times 10^{-7}.
\end{align}

In this mass range, additional constraints on the dark photon arise from stellar cooling. 
For masses lighter than twice the electron mass, cooling bounds from the Sun, horizontal branch stars, and red giants give a stringent constraint. 
In particular, in the range $10^{2}~\mr{eV} \lesssim m_{Z'} \lesssim 10^{4}~\mr{eV}$, only extremely small kinetic mixing values, $\epsilon \lesssim 10^{-14}$, are allowed~\cite{An:2013yfc, Redondo:2013lna}. 
In the same mass range, there are also cooling constraints from neutron stars~\cite{Hong:2020bxo}, although these are weaker than the cooling bounds from other types of stars. 
For further observational constraints on light dark photons, see Ref.~\cite{Caputo:2021eaa}.

\section{Summary}
\label{sec: summary}
In this work, we consider ADM  production associated with low-scale spontaneous leptogenesis,  
within a model that extends the type-I seesaw framework by incorporating a dark sector containing ADM.  In the low-scale spontaneous leptogenesis, the right-handed neutrinos remain in thermal equilibrium, and in the presence of a dynamically generated $CP$-violating background field (the majoron, in our case), the lepton asymmetry is produced through the inverse decays of SM leptons and the Higgs boson. In our setup, the right-handed neutrinos additionally have Yukawa interactions with the DM particle and a BSM scalar field. Consequently, inverse decays involving the DM and the BSM scalar simultaneously generate an asymmetry in the dark sector.

We found that, in the production of asymmetries, the Yukawa coupling $y_{\chi,1}$ between the lightest right-handed neutrino and the DM particle plays a crucial role,  governing both the evolution of the DM asymmetry and its final relic abundance. If this coupling is large, the asymmetry is rapidly produced up to its equilibrium value determined by the dynamical $CP$ phase, and then this value freezes out when the inverse decay processes become decoupled. In contrast, when the coupling is small, the asymmetry cannot reach its equilibrium value but is gradually produced and eventually freezes in. We refer to the former as the freeze-out scenario and to the latter as the freeze-in scenario. 
The boundary between these two regimes is determined by the size of the Yukawa interaction between the right-handed neutrinos and the SM leptons, $y_{N,1}$, and as the coupling deviates from this scale, the total amount of DM asymmetry produced decreases in both scenarios.

We then numerically solved the Boltzmann equations to compute the final relic asymmetry and the corresponding predicted DM mass in both the freeze-out and freeze-in scenarios.  
In particular, in the freeze-out scenario, the DM mass is typically predicted to be in the range  
$\mathcal{O}(0.1)~\mathrm{GeV} \lesssim m_{\chi} \lesssim \mathcal{O}(100)~\mathrm{GeV}$, as long as scattering can be neglected.
This suggests that our scenario might be tested by upcoming experiments such as direct detection searches. On the other hand, if the dark matter asymmetry does not reach its equilibrium value due to the weak coupling, the allowed mass range extends over a broader interval, 
$\mathcal{O}(0.1)~\mathrm{GeV} \lesssim m_{\chi} \lesssim \mathcal{O}(10)~\mr{TeV}$.

\section*{Acknowledgments}
We thank Koichi Hamaguchi and Natsumi Nagata for carefully reading the draft and for their helpful comments. This work was supported by JSPS KAKENHI (Grant Numbers 25KJ0022 [JW] and 24KJ0913 [HT]).

\appendix

\section{Boltzman equation}~\label{appendix: Boltzmann equation}
In this appendix, we summarize the Boltzmann equations used for numerical computation. 
Using the dimensionless parameter $z=M_1/T$, the Boltzmann equation for the lepton asymmetry, Eq.~\eqref{eq: Boltzmann equation for lepton sector}, and that for the DM asymmetry, Eq.~\eqref{eq: Boltzmann equation for dark sector}, are cast into the following forms:
\begin{align}\label{eq: dimensionless Boltzmann equation for lepton sector}
    \frac{dY_{\Delta l_{\alpha}}}{dz} 
    &= -z\frac{\langle \Gamma_{N_1\to l_{\alpha}H}\rangle}{H_1}Y_{N_1}^{\rm eq}\left(\frac{13}{7}\frac{Y_{\Delta l_{\alpha}}}{Y^{\rm eq}_{l_{\alpha}}}-\frac{ g_{N,1}^2}{2M_1^2}\frac{s(T)}{T}Y_\theta \right),\\
    \frac{dY_{\Delta \chi}}{dz} 
    &= -z\frac{\langle \Gamma_{N_1\to \chi \phi}\rangle}{H_1}Y_{N_1}^{\rm eq}\left(\frac{Y_{\Delta \chi}}{Y^{\rm eq}_\chi} -\frac{ g_{N,1}^2}{2M_1^2}\frac{s(T)}{T}Y_\theta\right),
\end{align}
where Eq.~\eqref{eq: conserved charge} is used.
For the SM sector, the chemical equilibrium condition
\begin{align}
  \mu_{H} = \frac{4}{7}\mu_{l_\alpha}
\end{align}
is used to relate $n_{\Delta l_\alpha}$ and $n_{\Delta H}$.

When all the SM interactions are in thermal equilibrium, i.e., for temperatures $T < 10^5~\mathrm{GeV}$, the equilibrium value of the lepton asymmetry is given by
\begin{align}
n_{\Delta l_{\alpha}}
	&= \frac{7}{13}\, n_{l_{\alpha}}^{\mathrm{eq}}\, \frac{\dot{\theta}}{T}, \\
	&= \frac{21}{26}\, \frac{\zeta(3)}{\pi^2}\, \dot{\theta}\, T^2,
\end{align}
where $n_{l_{\alpha}}^{\mathrm{eq}}$ denotes the equilibrium number density of leptons.  
In terms of the total lepton asymmetry, one obtains
\begin{align}
n_{\Delta L}
	&= \sum_{\alpha} \left(n_{\Delta l_{\alpha}}^{\mathrm{eq}} + n_{\Delta e_R^{\alpha}}^{\mathrm{eq}}\right), \\
	&= \frac{153}{52}\, \frac{\zeta(3)}{\pi^2}\, \dot{\theta}\, T^2.
\end{align}

For the dark sector, we assume that the asymmetry in $\phi$ is efficiently washed out, so that $n_{\Delta \phi}=0$.  
Then, the equilibrium value of the DM asymmetry is
\begin{align}
   n_{\Delta \chi}
      = \frac{9\,\zeta(3)}{12\pi^2}\, \dot{\theta}\, T^2.
\end{align}

\section{Annihilation of symmetric components}~\label{appendix: symmetric component}

In ADM models, the reduction of the symmetric component of DM is a central issue.  
In our scenario, the most minimal annihilation channel is the right-handed neutrino–mediated process $\chi \bar{\chi} \to \phi \phi^*$. However, due to the large mass of the right-handed neutrino and the $p$-wave suppression, this process is insufficient to reduce the symmetric component. A more efficient annihilation is provided by the additional channel into the hidden photon, $\chi \bar{\chi} \to Z' Z'$, introduced in the extended setup. Below, using quantitative estimates, we derive the difficulty inherent in the minimal model and determine how large coupling is required when a hidden photon is included.

Let $n_{\chi}$ and $n_{\bar{\chi}}$ denote the number densities of DM and anti-DM, respectively.  
After the DM asymmetry production has frozen out, the Boltzmann equation for the number density of the anti-DM that undergoes annihilation with the thermally-averaged cross section $\langle \sigma v \rangle$ reads
\begin{align}
\label{eq: boltzman equation for antiDM density}
\dot{n}_{\bar{\chi}} + 3 H n_{\bar{\chi}}
= - \braket{\sigma v} \bigl( n_{\chi} n_{\bar{\chi}} - n_{\chi}^{\mr{eq}} n_{\bar{\chi}}^{\mr{eq}} \bigr),
\end{align}
or, in its dimensionless form,
\begin{align}
\label{eq: boltzman equation for DM asymmetry}
\frac{dY_{\bar{\chi}}}{dx}
= - \lambda_{\chi}\, x^{-n-2} \bigl( Y_{\chi} Y_{\bar{\chi}} - Y_{\chi}^{\mr{eq}} Y_{\bar{\chi}}^{\mr{eq}} \bigr),
\end{align}
where
\begin{align}
& x = \frac{m_{\chi}}{T}, \qquad
\lambda_{\chi} := \left[ \frac{x\, s}{H(x)} \braket{\sigma v} \right]_{x = 1}.
\end{align}
Here,  $n = 0$ ($n = 1$) corresponds to the case of the s-wave (p-wave) process.
At late times when $Y_{\chi}^{\mr{eq}} Y_{\bar{\chi}}^{\mr{eq}}$ is negligible, the solution can be found analytically, under the assumption that $Y_{\Delta \chi} := Y_{\chi}(x) -Y_{\bar{\chi}}(x)$ is already frozen out:
\begin{align}
Y_{\bar{\chi}}(\infty) 
=\frac{Y_{\Delta \chi}}{e^{\lambda_\chi Y_{\Delta \chi} x_f^{-n-1}/(n+1)}\left[1+Y_{\Delta \chi} /Y_{\bar{\chi}}(x_f)\right]-1},
\label{eq: anti-DM symmetric abundance}
\end{align}
where $x_f \sim \mathcal{O}(1)$ denotes the value when $Y_{\chi}^{\mr{eq}} Y_{\bar{\chi}}^{\mr{eq}}$ starts to become negligible. 

For the symmetric component to be subdominant, $Y_{\Delta \chi} \gg Y_{\bar{\chi}}(\infty)$, or, written in a useful form,
\begin{align}
e^{\lambda_\chi Y_{\Delta \chi}} \gg 1,
\label{eq: condition for sufficient annihilation}
\end{align}
needs to be satisfied, where $x_f \sim \mathcal{O}(1)$ and $Y_{\bar{\chi}}(x_f)\lesssim Y_{\Delta \chi}$ are used to drop all the $\mathcal{O}(1)$ factors for simplicity.

Now consider the p-wave annihilation $\chi\bar\chi \to \phi\phi^*$, caused by the t-channel exchange of $N_1$.
The thermally-averaged annihilation cross-section for this process is given by
\begin{align}
\langle \sigma v \rangle \simeq \frac{|y_{\chi,1}|^4}{8\pi M_1^2} \frac{T}{m_\chi}.
\label{eq:pwave}
\end{align}
Using Eq.~\eqref{eq:pwave} and the ratio of the observed energy densities of DM and baryons, $\Omega_{\rm DM}/\Omega_B \simeq 5.4$~\cite{Planck:2018vyg}, the exponent in Eq.~\eqref{eq: condition for sufficient annihilation} is evaluated as
\begin{align}
    \lambda_\chi Y_{\Delta \chi} \simeq 10^{-12} \left( \frac{|y_{\chi,1}|}{10^{-3}}\right)^{4}\left( \frac{M_1}{10~{\rm TeV}}\right)^{-2},
\end{align}
showing that Eq.~\eqref{eq: condition for sufficient annihilation} is never satisfied in the parameter regions of interest.

The annihilation via gauge interaction, on the other hand, can be $s$-wave dominated. 
Assuming that $m_\chi > m_{Z^\prime}$, the process $\chi \bar{\chi} \to Z' Z'$ gives
\begin{align}
\langle \sigma v \rangle \simeq \frac{{g^\prime}^4}{8\pi m_{\chi}^2}.
\label{eq:swave}
\end{align}
In this case, the exponent in Eq.~\eqref{eq: condition for sufficient annihilation} becomes
\begin{align}
    \lambda_\chi Y_{\Delta \chi} \simeq \left( \frac{g^\prime}{10^{-2}}\right)^{4}\left( \frac{m_\chi}{1~{\rm GeV}}\right)^{-2},
\end{align}
implying that the symmetric component can be subdominant for 
\begin{align}\label{eq:cond. for sym comp of chi to annihilate}
    g^\prime \gtrsim 10^{-2} \left( \frac{m_\chi}{1~{\rm GeV}}\right)^{1/2}.
\end{align}
Given the perturbativity condition, $g^\prime < \sqrt{4\pi}$, this constraint leads to an upper bound on $m_\chi$:
\begin{align}
    m_\chi \lesssim 10^5~\text{GeV}.
\end{align}

\section{$\Delta L = 2$ scattering rate}~\label{app: delta 2 scattering rate}
In this appendix, we discuss the $\Delta L = 2$ scattering process $\chi \phi \to \bar{\chi}\,\phi^*$ considered in the main text.  
The interaction rate $\Gamma(\chi \phi \to \bar{\chi}\,\phi^*)$ is given by
\begin{align}
&\Gamma(\chi \phi \to \bar{\chi}\,\phi^*) \nonumber \\
	&\qquad
        \simeq {1 \over n_{\chi}^{\rm eq}}
		{T \over 32\pi^4}
		\int_0^{\infty} ds\, s^{3/2}\, K_1\!\left(\sqrt{s}/T\right)
		\sigma(\chi \phi \to \bar{\chi}\,\phi^*),
\label{appeq: scattering interaction rate s-channel}
\end{align}
where $\sigma(\chi \phi \to \bar{\chi}\,\phi^*)$ denotes the scattering cross section for this process, and we treat all particles except the right-handed neutrino as massless for simplicity.

This process contains both $s$- and $t$-channel contributions, but for simplicity, we first focus on the $s$-channel, whose scattering cross section is
\begin{align}
\label{appeq: scattering cross section s-channel}
\sigma(\chi \phi \to \bar{\chi}\,\phi^*)
	&= \frac{|y_{\chi,1}|^{4}}{32\pi}\, M_1^{2}\, |D_{N_1}|^{2},  \\
D_{N_1}
	&= \frac{1}{s - M_1^{2} + i M_1 \Gamma_1},
\end{align}
where
\begin{align}
\Gamma_1 &:= \Gamma_{N_1 \to \chi \phi} + \Gamma_{N_1 \to \bar{\chi}\,\phi^*} \nonumber \\
        &\qquad
        + \Gamma_{N_1 \to l_{\alpha} H} + \Gamma_{N_1 \to \bar l_{\alpha} H^*} 
\end{align}
is the total decay width. 

Equation~\eqref{appeq: scattering interaction rate s-channel} contains the on-shell resonance. Therefore, when formulating the Boltzmann equations that include both decays and scatterings, one must properly remove this resonance; otherwise, the result becomes an overestimate that double-counts the decay contribution~\cite{Buchmuller:2004nz, Davidson:2008bu}.
To remove this contribution, several subtraction prescriptions have been proposed. One approach treats the imaginary part of the
propagator as the on-shell contribution and regards the real part
\begin{align}
\label{appeq: PVS}
D_{N_1,\,{\rm off}}^{\rm PVS}(s) := \Re[D_{N_1}]
\end{align}
as the off-shell propagator whose square is to be used for computing the interaction rate; we refer to this as the principal value subtraction (PVS)~\cite{Luty:1992un, Plumacher:1996kc, Buchmuller:2000as, Buchmuller:2002jk, Buchmuller:2003gz, Pilaftsis:2003gt}.
Another approach considers the squared propagator and subtracts its imaginary part; we refer to this as the improved principal value subtraction (iPVS)~\cite{Cline:1993bd, Pilaftsis:2003gt, Giudice:2003jh}.

As pointed out in previous studies~\cite{Giudice:2003jh, Ala-Mattinen:2023rbm}, the PVS prescription subtracts only
half of the on-shell contribution. Indeed, one finds
\begin{align}
 |\Im[D_{N_1}]|^2 \;\to\; {\pi \over 2 M_1 \Gamma_1}\, \delta(s - M_1^{2}) ,
\end{align}
which contains only half of the full on-shell piece.  
On the other hand, the iPVS prescription suffers from a more fundamental problem: when the center-of-mass energy approaches the near on-shell region, the resulting reaction rate can become negative~\cite{Ala-Mattinen:2023rbm}.

Thus, we introduce the cut subtraction scheme, as in Ref.~\cite{Ala-Mattinen:2023rbm},  and use it to evaluate the interaction rate in Eq.~\eqref{appeq: scattering interaction rate s-channel}.  In addition, we derive an expression for estimating the accuracy of the subtraction, using a method different from that of Ref.~\cite{Ala-Mattinen:2023rbm}.

To evaluate Eq.~\eqref{appeq: scattering interaction rate s-channel}, we need to
consider the following integral
\begin{align}
J := \int_{0}^{\infty} ds\, G(s)\, F(s),
\end{align}
where
\begin{align}
G(s) &:= |D_{N_1}|^{2}
      = \frac{1}{(s - M_1^{2})^{2} + M_1^{2}\Gamma_1^{2}}, \\
F(s) &:= s^{3/2}\, K_{1}(\sqrt{s}/T).
\end{align}

We split the integration range into the near on-shell region $M_1^{2}-\kappa M_1 \Gamma_1 \le s \le M_1^{2}+\kappa M_1 \Gamma_1$ with $\kappa > 0$ and the remainder:
\begin{align}
\label{appeq: on-shell contribution}
J_{\rm on} &:= \int_{M_1^{2}-\kappa M_1 \Gamma_1}^{M_1^{2}+\kappa M_1 \Gamma_1} ds\, G(s)\, F(s), \\
\label{appeq: off-shell contribution}
J_{\rm off} &:= J - J_{\rm on}.
\end{align}
In the range $M_1^{2} - \kappa M_1 \Gamma_1 \le s \le M_1^{2} + \kappa M_1 \Gamma_1$, one may approximate $s \simeq M_1^{2}$ and expand $F(s)$ in a Taylor series:
\begin{align}
F(s)
       & \simeq F(M_1^{2})
		+ F_1(M_1^{2})(s - M_1^{2}) \\
	&+ \frac{1}{2} F_2(M_1^{2})(s - M_1^{2})^{2}
		+ \cdots ,
\end{align}
where $F_1(s) := \frac{d}{ds} F(s)$ and $F_2(s) := \frac{d^{2}}{ds^{2}} F(s)$.

Thus, for $\kappa \gg 1$, the quantity $J_{\rm on}$ can be decomposed into the
on-shell contribution and its correction terms:
\begin{align}
J_{\rm on}
	\simeq \frac{\pi}{M_1 \Gamma_1} F(M_1^{2})
		+ M_1 \Gamma_1 \left(\kappa - {\pi \over 2}\right) F_2(M_1^{2}).
\end{align}
We note that, in the regime of our interest where the Yukawa interactions of the right-handed neutrino are sufficiently small, one has $\Gamma_1/M_1 \ll 1$, and this expansion is well justified as long as $\Gamma_1 \kappa/M_1 \ll 1$.

In deriving this expression, we have used the integral formulas
\begin{align}
\int_{-\Delta}^{\Delta} \frac{dx}{x^{2} + a^{2}}
	&= \frac{2}{a} \arctan\!\left(\frac{\Delta}{a}\right), \\[4pt]
\int_{-\Delta}^{\Delta} \frac{x^{2}\, dx}{x^{2} + a^{2}}
	&= 2\Delta - 2a\, \arctan\!\left(\frac{\Delta}{a}\right),
\end{align}
together with the expansion valid for $\Delta \gg a$,
\begin{align}
\arctan\!\left(\frac{\Delta}{a}\right)
	\simeq \frac{\pi}{2} - \frac{a}{\Delta}.
\end{align}

The first term of $J_{\rm on}$ corresponds to the contribution obtained by approximating the squared propagator $G(s)$ by a delta function,
\begin{align}
G(s)
	\to \frac{\pi}{M_1 \Gamma_1} \delta(s-M_1^2),
\end{align}
which represents the on-shell resonance itself. Therefore, by separating the integration range as in Eqs.~\eqref{appeq: on-shell contribution} and \eqref{appeq: off-shell contribution} and evaluating only $J_{\mr{off}}$, one can isolate the off-shell contribution, although an ambiguity of the order of the second term in $J_{\mr{on}}$ remains.

The accuracy of this subtraction can be estimated by taking the ratio of the
second term in $J_{\rm on}$ to the first:
\begin{align}
R
	&= 
	\frac{
		M_1 \Gamma_1 \left(\kappa - \frac{\pi}{2}\right) F_2(M_1^{2})
	}{
		\frac{\pi}{M_1 \Gamma_1} F(M_1^{2})
	}
	\\[4pt]
	&\simeq 
	\frac{\Gamma_1^{2}}{M_1^{2}}
	\frac{\kappa}{\pi}
	\frac{z^{2}}{4}\,
	\bigl(1 - 4 K_{0}(z)\bigr),
\end{align}
which means that the correction is of order
$\mathcal{O}(z^2 \kappa {\Gamma_1^2 /M_1^2})$.

In the parameter region of our interest, the ratio $\Gamma_1^{2}/M_1^{2}$ is extremely small, and the subtraction is relevant only in the regime $z \lesssim 1$, where the contribution of the on-shell pole becomes significant. Hence, this expansion, together with the decomposition of $J$ in Eqs.~\eqref{appeq: on-shell contribution} and \eqref{appeq: off-shell contribution}, provides a subtraction with good accuracy. Moreover, unlike the iPVS prescription, it never yields a negative interaction rate.

\begin{figure}[t]
  \centering
  \includegraphics[width=0.45\textwidth]{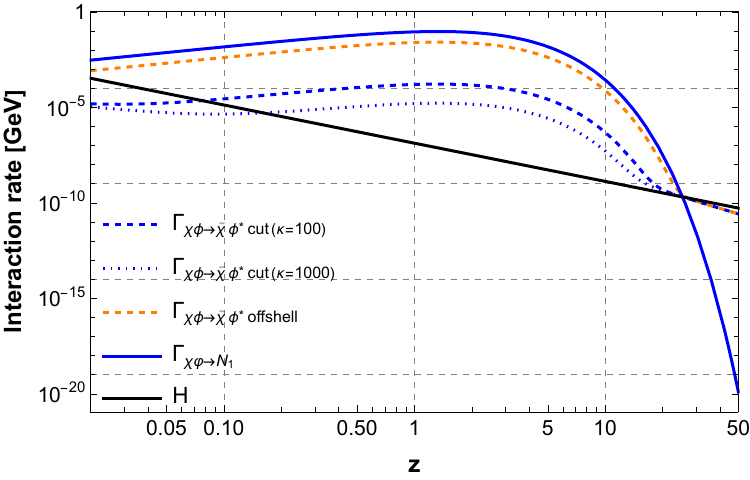}
  \caption{Comparison of the interaction rates for the $\Delta L = 2$ scattering process $\chi \phi \to \bar{\chi}\,\phi^*$ under different subtraction schemes. The curves correspond to the rates evaluated with cut subtraction with $\kappa = 100$ (blue dashed), $\kappa = 1000$ (blue dotted), and principal value subtraction (orange dashed). For reference, we also show the Hubble rate (black line) and the interaction rate of the inverse decay process $\chi \phi \to N_1$ (blue solid).}
  \label{fig: Comparsion of different subtraction}
\end{figure}

In Fig.~\ref{fig: Comparsion of different subtraction}, we compare the interaction rates obtained from different subtraction schemes at the benchmark point $M_1 = 300~\mathrm{TeV}$, $y_{N,1} = 2\times 10^{-5}$, and $|y_{\chi,1}| = y_{\chi,1}^{\rm sct}=7\times 10^{-3}$. 
The interaction rates shown in the figure include not only the $s$-channel contribution but also the $t$-channel one. 
The case with $\kappa = 100$ corresponds to the same choice as in Fig.~\ref{fig: interactionrate} of the main text, while $\kappa = 1000$ illustrates the behavior when $\kappa$ is increased by one order of magnitude. For reference, we also display the interaction rate obtained with the PVS prescription (Eq.~\eqref{appeq: PVS}).

The reason why the scattering interaction rate changes significantly when different values of $\kappa$ are chosen is that, in the parameter region of interest, there exists a hierarchy between the decay and scattering interaction rates. Thus, even if the subtraction of the on-shell contribution is accurate, the precision on the off-shell side is not guaranteed. Nevertheless, for the purpose of assessing whether the effects of scattering can be neglected in comparison with decays at $z \lesssim 1$, this uncertainty is practically irrelevant. Moreover, regardless of the subtraction scheme adopted, the behavior for $z > z_{\rm fo}^{D}$ agrees and becomes independent of the subtraction prescription. This is because, on the low-temperature side, namely at sufficiently large $z$, the on-shell resonance of the right-handed neutrino propagator is hardly accessible. Consequently, the result is insensitive to how the on-shell contribution is subtracted.

\bibliographystyle{utphysmod}
\bibliography{ref}

@article{Ibe:2015nfa,
    author = "Ibe, Masahiro and Kaneta, Kunio",
    title = "{Spontaneous thermal Leptogenesis via Majoron oscillation}",
    eprint = "1504.04125",
    archivePrefix = "arXiv",
    primaryClass = "hep-ph",
    reportNumber = "ICRR-REPORT-700-2015-1, IPMU15-0048",
    doi = "10.1103/PhysRevD.92.035019",
    journal = "Phys. Rev. D",
    volume = "92",
    number = "3",
    pages = "035019",
    year = "2015"
}

@article{Chun:2023eqc,
    author = "Chun, Eung Jin and Jung, Tae Hyun",
    title = "{Leptogenesis driven by a Majoron}",
    eprint = "2311.09005",
    archivePrefix = "arXiv",
    primaryClass = "hep-ph",
    reportNumber = "CTPU-PTC-23-48",
    doi = "10.1103/PhysRevD.109.095004",
    journal = "Phys. Rev. D",
    volume = "109",
    number = "9",
    pages = "095004",
    year = "2024"
}

@article{Wada:2024cbe,
    author = "Wada, Juntaro",
    title = "{Majoron-driven leptogenesis in gauged U(1)L{\ensuremath{\mu}}-L{\ensuremath{\tau}} model}",
    eprint = "2404.10283",
    archivePrefix = "arXiv",
    primaryClass = "hep-ph",
    doi = "10.1103/PhysRevD.110.103510",
    journal = "Phys. Rev. D",
    volume = "110",
    number = "10",
    pages = "103510",
    year = "2024"
}

@article{Planck:2018vyg,
    author = "Aghanim, N. and others",
    collaboration = "Planck",
    title = "{Planck 2018 results. VI. Cosmological parameters}",
    eprint = "1807.06209",
    archivePrefix = "arXiv",
    primaryClass = "astro-ph.CO",
    doi = "10.1051/0004-6361/201833910",
    journal = "Astron. Astrophys.",
    volume = "641",
    pages = "A6",
    year = "2020",
    note = "[Erratum: Astron.Astrophys. 652, C4 (2021)]"
}

@article{Nussinov:1985xr,
    author = "Nussinov, S.",
    title = "{TECHNOCOSMOLOGY: COULD A TECHNIBARYON EXCESS PROVIDE A 'NATURAL' MISSING MASS CANDIDATE?}",
    reportNumber = "CLNS-85/703",
    doi = "10.1016/0370-2693(85)90689-6",
    journal = "Phys. Lett. B",
    volume = "165",
    pages = "55--58",
    year = "1985"
}

@article{Kaplan:1991ah,
    author = "Kaplan, David B.",
    title = "{A Single explanation for both the baryon and dark matter densities}",
    reportNumber = "UCSD-PTH-91-22",
    doi = "10.1103/PhysRevLett.68.741",
    journal = "Phys. Rev. Lett.",
    volume = "68",
    pages = "741--743",
    year = "1992"
}

@article{Minkowski:1977sc,
    author = "Minkowski, Peter",
    title = "{$\mu \to e\gamma$ at a Rate of One Out of $10^{9}$ Muon Decays?}",
    reportNumber = "Print-77-0182 (BERN)",
    doi = "10.1016/0370-2693(77)90435-X",
    journal = "Phys. Lett. B",
    volume = "67",
    pages = "421--428",
    year = "1977"
}

@article{Gell-Mann:1979vob,
    author = "Gell-Mann, Murray and Ramond, Pierre and Slansky, Richard",
    title = "{Complex Spinors and Unified Theories}",
    eprint = "1306.4669",
    archivePrefix = "arXiv",
    primaryClass = "hep-th",
    reportNumber = "PRINT-80-0576",
    journal = "Conf. Proc. C",
    volume = "790927",
    pages = "315--321",
    year = "1979"
}

@article{Mohapatra:1979ia,
    author = "Mohapatra, Rabindra N. and Senjanovic, Goran",
    title = "{Neutrino Mass and Spontaneous Parity Nonconservation}",
    reportNumber = "MDDP-TR-80-060, MDDP-PP-80-105, CCNY-HEP-79-10",
    doi = "10.1103/PhysRevLett.44.912",
    journal = "Phys. Rev. Lett.",
    volume = "44",
    pages = "912",
    year = "1980"
}

@article{Falkowski:2011xh,
    author = "Falkowski, Adam and Ruderman, Joshua T. and Volansky, Tomer",
    title = "{Asymmetric Dark Matter from Leptogenesis}",
    eprint = "1101.4936",
    archivePrefix = "arXiv",
    primaryClass = "hep-ph",
    reportNumber = "LPT-ORSAY-11-09",
    doi = "10.1007/JHEP05(2011)106",
    journal = "JHEP",
    volume = "05",
    pages = "106",
    year = "2011"
}

@article{Cohen:1987vi,
    author = "Cohen, Andrew G. and Kaplan, David B.",
    title = "{Thermodynamic Generation of the Baryon Asymmetry}",
    reportNumber = "HUTP-87/A061",
    doi = "10.1016/0370-2693(87)91369-4",
    journal = "Phys. Lett. B",
    volume = "199",
    pages = "251--258",
    year = "1987"
}

@article{Cohen:1988kt,
    author = "Cohen, Andrew G. and Kaplan, David B.",
    title = "{SPONTANEOUS BARYOGENESIS}",
    reportNumber = "HUTP-88/A016",
    doi = "10.1016/0550-3213(88)90134-4",
    journal = "Nucl. Phys. B",
    volume = "308",
    pages = "913--928",
    year = "1988"
}

@article{Buchmuller:2004nz,
    author = "Buchmuller, W. and Di Bari, P. and Plumacher, M.",
    title = "{Leptogenesis for pedestrians}",
    eprint = "hep-ph/0401240",
    archivePrefix = "arXiv",
    reportNumber = "DESY-03-100, UAB-FT-551, CERN-TH-2003-199",
    doi = "10.1016/j.aop.2004.02.003",
    journal = "Annals Phys.",
    volume = "315",
    pages = "305--351",
    year = "2005"
}

@article{Banks:2010zn,
    author = "Banks, Tom and Seiberg, Nathan",
    title = "{Symmetries and Strings in Field Theory and Gravity}",
    eprint = "1011.5120",
    archivePrefix = "arXiv",
    primaryClass = "hep-th",
    doi = "10.1103/PhysRevD.83.084019",
    journal = "Phys. Rev. D",
    volume = "83",
    pages = "084019",
    year = "2011"
}

@article{Witten:2017hdv,
    author = "Witten, Edward",
    title = "{Symmetry and Emergence}",
    eprint = "1710.01791",
    archivePrefix = "arXiv",
    primaryClass = "hep-th",
    doi = "10.1038/nphys4348",
    journal = "Nature Phys.",
    volume = "14",
    number = "2",
    pages = "116--119",
    year = "2018"
}

@article{Harlow:2018jwu,
    author = "Harlow, Daniel and Ooguri, Hirosi",
    title = "{Constraints on Symmetries from Holography}",
    eprint = "1810.05337",
    archivePrefix = "arXiv",
    primaryClass = "hep-th",
    doi = "10.1103/PhysRevLett.122.191601",
    journal = "Phys. Rev. Lett.",
    volume = "122",
    number = "19",
    pages = "191601",
    year = "2019"
}

@article{Harlow:2018tng,
    author = "Harlow, Daniel and Ooguri, Hirosi",
    title = "{Symmetries in quantum field theory and quantum gravity}",
    eprint = "1810.05338",
    archivePrefix = "arXiv",
    primaryClass = "hep-th",
    doi = "10.1007/s00220-021-04040-y",
    journal = "Commun. Math. Phys.",
    volume = "383",
    number = "3",
    pages = "1669--1804",
    year = "2021"
}

@article{Berbig:2023uzs,
    author = "Berbig, Maximilian",
    title = "{Diraxiogenesis}",
    eprint = "2307.14121",
    archivePrefix = "arXiv",
    primaryClass = "hep-ph",
    doi = "10.1007/JHEP01(2024)061",
    journal = "JHEP",
    volume = "01",
    pages = "061",
    year = "2024"
}

@article{Kusenko:2014uta,
    author = "Kusenko, Alexander and Schmitz, Kai and Yanagida, Tsutomu T.",
    title = "{Leptogenesis via Axion Oscillations after Inflation}",
    eprint = "1412.2043",
    archivePrefix = "arXiv",
    primaryClass = "hep-ph",
    reportNumber = "IPMU-14-0352",
    doi = "10.1103/PhysRevLett.115.011302",
    journal = "Phys. Rev. Lett.",
    volume = "115",
    number = "1",
    pages = "011302",
    year = "2015"
}

@article{Domcke:2020kcp,
    author = "Domcke, Valerie and Ema, Yohei and Mukaida, Kyohei and Yamada, Masaki",
    title = "{Spontaneous Baryogenesis from Axions with Generic Couplings}",
    eprint = "2006.03148",
    archivePrefix = "arXiv",
    primaryClass = "hep-ph",
    reportNumber = "DESY 20-100, DESY-20-100, CERN-TH-2020-088, TU-1102",
    doi = "10.1007/JHEP08(2020)096",
    journal = "JHEP",
    volume = "08",
    pages = "096",
    year = "2020"
}

@article{Co:2020jtv,
    author = "Co, Raymond T. and Fernandez, Nicolas and Ghalsasi, Akshay and Hall, Lawrence J. and Harigaya, Keisuke",
    title = "{Lepto-Axiogenesis}",
    eprint = "2006.05687",
    archivePrefix = "arXiv",
    primaryClass = "hep-ph",
    doi = "10.1007/JHEP03(2021)017",
    journal = "JHEP",
    volume = "03",
    pages = "017",
    year = "2021"
}

@article{Domcke:2020quw,
    author = "Domcke, Valerie and Kamada, Kohei and Mukaida, Kyohei and Schmitz, Kai and Yamada, Masaki",
    title = "{Wash-In Leptogenesis}",
    eprint = "2011.09347",
    archivePrefix = "arXiv",
    primaryClass = "hep-ph",
    reportNumber = "CERN-TH-2020-196, RESCEU-22/20, DESY-20-202, TU-1112",
    doi = "10.1103/PhysRevLett.126.201802",
    journal = "Phys. Rev. Lett.",
    volume = "126",
    number = "20",
    pages = "201802",
    year = "2021"
}

@article{Chao:2023ojl,
    author = "Chao, Wei and Peng, Ying-Quan",
    title = "{Majorana Majoron and the Baryon Asymmetry of the Universe}",
    eprint = "2311.06469",
    archivePrefix = "arXiv",
    primaryClass = "hep-ph",
    month = "11",
    year = "2023"
}

@article{Barnes:2024jap,
    author = "Barnes, Patrick and Co, Raymond T. and Harigaya, Keisuke and Pierce, Aaron",
    title = "{Lepto-axiogenesis with light right-handed neutrinos}",
    eprint = "2402.10263",
    archivePrefix = "arXiv",
    primaryClass = "hep-ph",
    reportNumber = "LCTP-24-02, UMN-TH-4307/23, FTPI-MINN-23-27",
    doi = "10.1007/JHEP08(2025)004",
    journal = "JHEP",
    volume = "08",
    pages = "004",
    year = "2025"
}

@article{Kuzmin:1985mm,
    author = "Kuzmin, V. A. and Rubakov, V. A. and Shaposhnikov, M. E.",
    title = "{On the Anomalous Electroweak Baryon Number Nonconservation in the Early Universe}",
    reportNumber = "IC/85/8",
    doi = "10.1016/0370-2693(85)91028-7",
    journal = "Phys. Lett. B",
    volume = "155",
    pages = "36",
    year = "1985"
}

@article{Co:2020xlh,
    author = "Co, Raymond T. and Hall, Lawrence J. and Harigaya, Keisuke",
    title = "{Predictions for Axion Couplings from ALP Cogenesis}",
    eprint = "2006.04809",
    archivePrefix = "arXiv",
    primaryClass = "hep-ph",
    reportNumber = "LCTP-20-11",
    doi = "10.1007/JHEP01(2021)172",
    journal = "JHEP",
    volume = "01",
    pages = "172",
    year = "2021"
}

@article{DOnofrio:2014rug,
    author = "D'Onofrio, Michela and Rummukainen, Kari and Tranberg, Anders",
    title = "{Sphaleron Rate in the Minimal Standard Model}",
    eprint = "1404.3565",
    archivePrefix = "arXiv",
    primaryClass = "hep-ph",
    doi = "10.1103/PhysRevLett.113.141602",
    journal = "Phys. Rev. Lett.",
    volume = "113",
    number = "14",
    pages = "141602",
    year = "2014"
}

@article{Davidson:2002qv,
    author = "Davidson, Sacha and Ibarra, Alejandro",
    title = "{A Lower bound on the right-handed neutrino mass from leptogenesis}",
    eprint = "hep-ph/0202239",
    archivePrefix = "arXiv",
    reportNumber = "OUTP-02-10P, IPPP-02-16, DCPT-02-32",
    doi = "10.1016/S0370-2693(02)01735-5",
    journal = "Phys. Lett. B",
    volume = "535",
    pages = "25--32",
    year = "2002"
}

@article{Co:2019jts,
    author = "Co, Raymond T. and Hall, Lawrence J. and Harigaya, Keisuke",
    title = "{Axion Kinetic Misalignment Mechanism}",
    eprint = "1910.14152",
    archivePrefix = "arXiv",
    primaryClass = "hep-ph",
    reportNumber = "LCTP-19-28",
    doi = "10.1103/PhysRevLett.124.251802",
    journal = "Phys. Rev. Lett.",
    volume = "124",
    number = "25",
    pages = "251802",
    year = "2020"
}

@article{Chang:2019tvx,
    author = "Chang, Chia-Feng and Cui, Yanou",
    title = "{New Perspectives on Axion Misalignment Mechanism}",
    eprint = "1911.11885",
    archivePrefix = "arXiv",
    primaryClass = "hep-ph",
    doi = "10.1103/PhysRevD.102.015003",
    journal = "Phys. Rev. D",
    volume = "102",
    number = "1",
    pages = "015003",
    year = "2020"
}

@article{Pilaftsis:1997jf,
    author = "Pilaftsis, Apostolos",
    title = "{CP violation and baryogenesis due to heavy Majorana neutrinos}",
    eprint = "hep-ph/9707235",
    archivePrefix = "arXiv",
    reportNumber = "MPI-PHT-97-30",
    doi = "10.1103/PhysRevD.56.5431",
    journal = "Phys. Rev. D",
    volume = "56",
    pages = "5431--5451",
    year = "1997"
}

@article{Pilaftsis:1997dr,
    author = "Pilaftsis, Apostolos",
    title = "{Resonant CP violation induced by particle mixing in transition amplitudes}",
    eprint = "hep-ph/9702393",
    archivePrefix = "arXiv",
    reportNumber = "MPI-PHT-97-002",
    doi = "10.1016/S0550-3213(97)00469-0",
    journal = "Nucl. Phys. B",
    volume = "504",
    pages = "61--107",
    year = "1997"
}

@article{Pilaftsis:2003gt,
    author = "Pilaftsis, Apostolos and Underwood, Thomas E. J.",
    title = "{Resonant leptogenesis}",
    eprint = "hep-ph/0309342",
    archivePrefix = "arXiv",
    reportNumber = "MC-TH-2003-09",
    doi = "10.1016/j.nuclphysb.2004.05.029",
    journal = "Nucl. Phys. B",
    volume = "692",
    pages = "303--345",
    year = "2004"
}

@article{Hambye:2003rt,
    author = "Hambye, Thomas and Lin, Yin and Notari, Alessio and Papucci, Michele and Strumia, Alessandro",
    title = "{Constraints on neutrino masses from leptogenesis models}",
    eprint = "hep-ph/0312203",
    archivePrefix = "arXiv",
    reportNumber = "IFUP-TH-2003-48",
    doi = "10.1016/j.nuclphysb.2004.06.027",
    journal = "Nucl. Phys. B",
    volume = "695",
    pages = "169--191",
    year = "2004"
}

@article{Blanchet:2008pw,
    author = "Blanchet, Steve and Di Bari, Pasquale",
    title = "{New aspects of leptogenesis bounds}",
    eprint = "0807.0743",
    archivePrefix = "arXiv",
    primaryClass = "hep-ph",
    doi = "10.1016/j.nuclphysb.2008.08.026",
    journal = "Nucl. Phys. B",
    volume = "807",
    pages = "155--187",
    year = "2009"
}

@article{Buckley:2011kk,
    author = "Buckley, Matthew R.",
    title = "{Asymmetric Dark Matter and Effective Operators}",
    eprint = "1104.1429",
    archivePrefix = "arXiv",
    primaryClass = "hep-ph",
    reportNumber = "FERMILAB-PUB-11-168-A",
    doi = "10.1103/PhysRevD.84.043510",
    journal = "Phys. Rev. D",
    volume = "84",
    pages = "043510",
    year = "2011"
}

@article{March-Russell:2012elz,
    author = "March-Russell, John and Unwin, James and West, Stephen M.",
    title = "{Closing in on Asymmetric Dark Matter I: Model independent limits for interactions with quarks}",
    eprint = "1203.4854",
    archivePrefix = "arXiv",
    primaryClass = "hep-ph",
    reportNumber = "OUTP-12-01P",
    doi = "10.1007/JHEP08(2012)029",
    journal = "JHEP",
    volume = "08",
    pages = "029",
    year = "2012"
}

@article{Ruderman:2009tj,
    author = "Ruderman, Joshua T. and Volansky, Tomer",
    title = "{Decaying into the Hidden Sector}",
    eprint = "0908.1570",
    archivePrefix = "arXiv",
    primaryClass = "hep-ph",
    doi = "10.1007/JHEP02(2010)024",
    journal = "JHEP",
    volume = "02",
    pages = "024",
    year = "2010"
}

@article{Hooper:2004dc,
    author = "Hooper, Dan and March-Russell, John and West, Stephen M.",
    title = "{Asymmetric sneutrino dark matter and the Omega(b) / Omega(DM) puzzle}",
    eprint = "hep-ph/0410114",
    archivePrefix = "arXiv",
    reportNumber = "OUTP-04-17P",
    doi = "10.1016/j.physletb.2004.11.047",
    journal = "Phys. Lett. B",
    volume = "605",
    pages = "228--236",
    year = "2005"
}

@article{Kaplan:2009ag,
    author = "Kaplan, David E. and Luty, Markus A. and Zurek, Kathryn M.",
    title = "{Asymmetric Dark Matter}",
    eprint = "0901.4117",
    archivePrefix = "arXiv",
    primaryClass = "hep-ph",
    reportNumber = "FERMILAB-PUB-09-345-A-T",
    doi = "10.1103/PhysRevD.79.115016",
    journal = "Phys. Rev. D",
    volume = "79",
    pages = "115016",
    year = "2009"
}

@article{Cohen:2009fz,
    author = "Cohen, Timothy and Zurek, Kathryn M.",
    title = "{Leptophilic Dark Matter from the Lepton Asymmetry}",
    eprint = "0909.2035",
    archivePrefix = "arXiv",
    primaryClass = "hep-ph",
    reportNumber = "FERMILAB-PUB-09-426-T, MCTP-09-45",
    doi = "10.1103/PhysRevLett.104.101301",
    journal = "Phys. Rev. Lett.",
    volume = "104",
    pages = "101301",
    year = "2010"
}

@article{Blennow:2010qp,
    author = "Blennow, Mattias and Dasgupta, Basudeb and Fernandez-Martinez, Enrique and Rius, Nuria",
    title = "{Aidnogenesis via Leptogenesis and Dark Sphalerons}",
    eprint = "1009.3159",
    archivePrefix = "arXiv",
    primaryClass = "hep-ph",
    reportNumber = "MPP-2010-125, IFIC-10-32, FTUV-10-0909",
    doi = "10.1007/JHEP03(2011)014",
    journal = "JHEP",
    volume = "03",
    pages = "014",
    year = "2011"
}

@article{Haba:2011uz,
    author = "Haba, Naoyuki and Matsumoto, Shigeki and Sato, Ryosuke",
    title = "{Sneutrino Inflation with Asymmetric Dark Matter}",
    eprint = "1101.5679",
    archivePrefix = "arXiv",
    primaryClass = "hep-ph",
    reportNumber = "IPMU-11-0011, OU-HET-691-2010, UT-11-04",
    doi = "10.1103/PhysRevD.84.055016",
    journal = "Phys. Rev. D",
    volume = "84",
    pages = "055016",
    year = "2011"
}

@article{Servant:2013uwa,
    author = "Servant, Geraldine and Tulin, Sean",
    title = "{Baryogenesis and Dark Matter through a Higgs Asymmetry}",
    eprint = "1304.3464",
    archivePrefix = "arXiv",
    primaryClass = "hep-ph",
    reportNumber = "CERN-PH-TH-2013-068",
    doi = "10.1103/PhysRevLett.111.151601",
    journal = "Phys. Rev. Lett.",
    volume = "111",
    number = "15",
    pages = "151601",
    year = "2013"
}

@article{Foot:2003jt,
    author = "Foot, Robert and Volkas, R. R.",
    title = "{Was ordinary matter synthesized from mirror matter? An Attempt to explain why Omega(Baryon) approximately equal to 0.2 Omega(Dark)}",
    eprint = "hep-ph/0304261",
    archivePrefix = "arXiv",
    doi = "10.1103/PhysRevD.68.021304",
    journal = "Phys. Rev. D",
    volume = "68",
    pages = "021304",
    year = "2003"
}

@article{Foot:2004pq,
    author = "Foot, Robert and Volkas, R. R.",
    title = "{Explaining Omega(Baryon) approximately 0.2 Omega(Dark) through the synthesis of ordinary matter from mirror matter: A More general analysis}",
    eprint = "hep-ph/0402267",
    archivePrefix = "arXiv",
    doi = "10.1103/PhysRevD.69.123510",
    journal = "Phys. Rev. D",
    volume = "69",
    pages = "123510",
    year = "2004"
}

@article{Shelton:2010ta,
    author = "Shelton, Jessie and Zurek, Kathryn M.",
    title = "{Darkogenesis: A baryon asymmetry from the dark matter sector}",
    eprint = "1008.1997",
    archivePrefix = "arXiv",
    primaryClass = "hep-ph",
    reportNumber = "MCTP-10-28",
    doi = "10.1103/PhysRevD.82.123512",
    journal = "Phys. Rev. D",
    volume = "82",
    pages = "123512",
    year = "2010"
}

@article{Haba:2010bm,
    author = "Haba, N. and Matsumoto, S.",
    title = "{Baryogenesis from Dark Sector}",
    eprint = "1008.2487",
    archivePrefix = "arXiv",
    primaryClass = "hep-ph",
    doi = "10.1143/PTP.125.1311",
    journal = "Prog. Theor. Phys.",
    volume = "125",
    pages = "1311--1316",
    year = "2011"
}

@article{Buckley:2010ui,
    author = "Buckley, Matthew R. and Randall, Lisa",
    title = "{Xogenesis}",
    eprint = "1009.0270",
    archivePrefix = "arXiv",
    primaryClass = "hep-ph",
    reportNumber = "FERMILAB-PUB-10-349-A",
    doi = "10.1007/JHEP09(2011)009",
    journal = "JHEP",
    volume = "09",
    pages = "009",
    year = "2011"
}

@article{Dutta:2010va,
    author = "Dutta, Bhaskar and Kumar, Jason",
    title = "{Asymmetric Dark Matter from Hidden Sector Baryogenesis}",
    eprint = "1012.1341",
    archivePrefix = "arXiv",
    primaryClass = "hep-ph",
    reportNumber = "MIFPA-10-52, UH-511-1156-2010",
    doi = "10.1016/j.physletb.2011.04.036",
    journal = "Phys. Lett. B",
    volume = "699",
    pages = "364--367",
    year = "2011"
}

@article{Davoudiasl:2010am,
    author = "Davoudiasl, Hooman and Morrissey, David E. and Sigurdson, Kris and Tulin, Sean",
    title = "{Hylogenesis: A Unified Origin for Baryonic Visible Matter and Antibaryonic Dark Matter}",
    eprint = "1008.2399",
    archivePrefix = "arXiv",
    primaryClass = "hep-ph",
    doi = "10.1103/PhysRevLett.105.211304",
    journal = "Phys. Rev. Lett.",
    volume = "105",
    pages = "211304",
    year = "2010"
}

@article{Feng:2013wn,
    author = "Feng, Wan-Zhe and Mazumdar, Anupam and Nath, Pran",
    title = "{Baryogenesis from dark matter}",
    eprint = "1302.0012",
    archivePrefix = "arXiv",
    primaryClass = "hep-ph",
    doi = "10.1103/PhysRevD.88.036014",
    journal = "Phys. Rev. D",
    volume = "88",
    number = "3",
    pages = "036014",
    year = "2013"
}

@article{Falkowski:2017uya,
    author = "Falkowski, Adam and Kuflik, Eric and Levi, Noam and Volansky, Tomer",
    title = "{Light Dark Matter from Leptogenesis}",
    eprint = "1712.07652",
    archivePrefix = "arXiv",
    primaryClass = "hep-ph",
    doi = "10.1103/PhysRevD.99.015022",
    journal = "Phys. Rev. D",
    volume = "99",
    number = "1",
    pages = "015022",
    year = "2019"
}

@article{Fukugita:1986hr,
    author = "Fukugita, M. and Yanagida, T.",
    title = "{Baryogenesis Without Grand Unification}",
    reportNumber = "RIFP-641",
    doi = "10.1016/0370-2693(86)91126-3",
    journal = "Phys. Lett. B",
    volume = "174",
    pages = "45--47",
    year = "1986"
}

@article{March-Russell:2011ang,
    author = "March-Russell, John and McCullough, Matthew",
    title = "{Asymmetric Dark Matter via Spontaneous Co-Genesis}",
    eprint = "1106.4319",
    archivePrefix = "arXiv",
    primaryClass = "hep-ph",
    reportNumber = "OUTP-11-42P",
    doi = "10.1088/1475-7516/2012/03/019",
    journal = "JCAP",
    volume = "03",
    pages = "019",
    year = "2012"
}

@article{Kamada:2012ht,
    author = "Kamada, Kohei and Yamaguchi, Masahide",
    title = "{Asymmetric Dark Matter from Spontaneous Cogenesis in the Supersymmetric Standard Model}",
    eprint = "1201.2636",
    archivePrefix = "arXiv",
    primaryClass = "hep-ph",
    reportNumber = "DESY-12-004",
    doi = "10.1103/PhysRevD.85.103530",
    journal = "Phys. Rev. D",
    volume = "85",
    pages = "103530",
    year = "2012"
}

@article{Cheung:2011if,
    author = "Cheung, Clifford and Zurek, Kathryn M.",
    title = "{Affleck-Dine Cogenesis}",
    eprint = "1105.4612",
    archivePrefix = "arXiv",
    primaryClass = "hep-ph",
    doi = "10.1103/PhysRevD.84.035007",
    journal = "Phys. Rev. D",
    volume = "84",
    pages = "035007",
    year = "2011"
}

@article{vonHarling:2012yn,
    author = "von Harling, Benedict and Petraki, Kalliopi and Volkas, Raymond R.",
    title = "{Affleck-Dine dynamics and the dark sector of pangenesis}",
    eprint = "1201.2200",
    archivePrefix = "arXiv",
    primaryClass = "hep-ph",
    doi = "10.1088/1475-7516/2012/05/021",
    journal = "JCAP",
    volume = "05",
    pages = "021",
    year = "2012"
}

@article{Borah:2022qln,
    author = "Borah, Debasish and Jyoti Das, Suruj and Okada, Nobuchika",
    title = "{Affleck-Dine cogenesis of baryon and dark matter}",
    eprint = "2212.04516",
    archivePrefix = "arXiv",
    primaryClass = "hep-ph",
    doi = "10.1007/JHEP05(2023)004",
    journal = "JHEP",
    volume = "05",
    pages = "004",
    year = "2023"
}

@article{Yanagida:1979as,
    author = "Yanagida, Tsutomu",
    editor = "Sawada, Osamu and Sugamoto, Akio",
    title = "{Horizontal gauge symmetry and masses of neutrinos}",
    reportNumber = "KEK-79-18-95",
    journal = "Conf. Proc. C",
    volume = "7902131",
    pages = "95--99",
    year = "1979"
}

@article{Luty:1992un,
    author = "Luty, M. A.",
    title = "{Baryogenesis via leptogenesis}",
    doi = "10.1103/PhysRevD.45.455",
    journal = "Phys. Rev. D",
    volume = "45",
    pages = "455--465",
    year = "1992"
}

@article{Plumacher:1996kc,
    author = "Plumacher, Michael",
    title = "{Baryogenesis and lepton number violation}",
    eprint = "hep-ph/9604229",
    archivePrefix = "arXiv",
    reportNumber = "DESY-96-052",
    doi = "10.1007/s002880050418",
    journal = "Z. Phys. C",
    volume = "74",
    pages = "549--559",
    year = "1997"
}

@article{Buchmuller:2000as,
    author = "Buchmuller, W. and Plumacher, M.",
    title = "{Neutrino masses and the baryon asymmetry}",
    eprint = "hep-ph/0007176",
    archivePrefix = "arXiv",
    reportNumber = "DESY-99-187, UPR-892-T",
    doi = "10.1016/S0217-751X(00)00293-5",
    journal = "Int. J. Mod. Phys. A",
    volume = "15",
    pages = "5047--5086",
    year = "2000"
}

@article{Buchmuller:2002jk,
    author = "Buchmuller, W. and Di Bari, P. and Plumacher, M.",
    title = "{A Bound on neutrino masses from baryogenesis}",
    eprint = "hep-ph/0209301",
    archivePrefix = "arXiv",
    reportNumber = "DESY-02-107, OUTP-02-30-P",
    doi = "10.1016/S0370-2693(02)02758-2",
    journal = "Phys. Lett. B",
    volume = "547",
    pages = "128--132",
    year = "2002"
}

@article{Buchmuller:2003gz,
    author = "Buchmuller, W. and Di Bari, P. and Plumacher, M.",
    title = "{The Neutrino mass window for baryogenesis}",
    eprint = "hep-ph/0302092",
    archivePrefix = "arXiv",
    reportNumber = "DESY-03-001, UAB-FT-540, CERN-TH-2003-016",
    doi = "10.1016/S0550-3213(03)00449-8",
    journal = "Nucl. Phys. B",
    volume = "665",
    pages = "445--468",
    year = "2003"
}

@article{Giudice:2003jh,
    author = "Giudice, G. F. and Notari, A. and Raidal, M. and Riotto, A. and Strumia, A.",
    title = "{Towards a complete theory of thermal leptogenesis in the SM and MSSM}",
    eprint = "hep-ph/0310123",
    archivePrefix = "arXiv",
    reportNumber = "IFUP-TH-2003-37, CERN-TH-2003-240",
    doi = "10.1016/j.nuclphysb.2004.02.019",
    journal = "Nucl. Phys. B",
    volume = "685",
    pages = "89--149",
    year = "2004"
}

@article{Cline:1993bd,
    author = "Cline, James M. and Kainulainen, Kimmo and Olive, Keith A.",
    title = "{Protecting the primordial baryon asymmetry from erasure by sphalerons}",
    eprint = "hep-ph/9401208",
    archivePrefix = "arXiv",
    reportNumber = "UMN-TH-1213-93, TPI-MINN-93-48-T",
    doi = "10.1103/PhysRevD.49.6394",
    journal = "Phys. Rev. D",
    volume = "49",
    pages = "6394--6409",
    year = "1994"
}

@article{Ala-Mattinen:2023rbm,
    author = "Ala-Mattinen, Kalle and Heikinheimo, Matti and Tuominen, Kimmo and Kainulainen, Kimmo",
    title = "{Anatomy of real intermediate state-subtraction scheme}",
    eprint = "2309.16615",
    archivePrefix = "arXiv",
    primaryClass = "hep-ph",
    reportNumber = "HIP-2023-14/TH",
    doi = "10.1103/PhysRevD.108.096034",
    journal = "Phys. Rev. D",
    volume = "108",
    number = "9",
    pages = "096034",
    year = "2023"
}

@article{Chiba:2003vp,
    author = "Chiba, Takeshi and Takahashi, Fuminobu and Yamaguchi, Masahide",
    title = "{Baryogenesis in a flat direction with neither baryon nor lepton charge}",
    eprint = "hep-ph/0304102",
    archivePrefix = "arXiv",
    reportNumber = "KUNS-1834, BROWN-HET-1351, RESCEU-47-03",
    doi = "10.1103/PhysRevLett.92.011301",
    journal = "Phys. Rev. Lett.",
    volume = "92",
    pages = "011301",
    year = "2004",
    note = "[Erratum: Phys.Rev.Lett. 114, 209901 (2015)]"
}

@article{Kusenko:2014lra,
    author = "Kusenko, Alexander and Pearce, Lauren and Yang, Louis",
    title = "{Postinflationary Higgs relaxation and the origin of matter-antimatter asymmetry}",
    eprint = "1410.0722",
    archivePrefix = "arXiv",
    primaryClass = "hep-ph",
    reportNumber = "FTPI-MINN-14-32, IPMU15-0005",
    doi = "10.1103/PhysRevLett.114.061302",
    journal = "Phys. Rev. Lett.",
    volume = "114",
    number = "6",
    pages = "061302",
    year = "2015"
}

@article{Pearce:2015nga,
    author = "Pearce, Lauren and Yang, Louis and Kusenko, Alexander and Peloso, Marco",
    title = "{Leptogenesis via neutrino production during Higgs condensate relaxation}",
    eprint = "1505.02461",
    archivePrefix = "arXiv",
    primaryClass = "hep-ph",
    doi = "10.1103/PhysRevD.92.023509",
    journal = "Phys. Rev. D",
    volume = "92",
    number = "2",
    pages = "023509",
    year = "2015"
}

@article{Nardi:2006fx,
    author = "Nardi, Enrico and Nir, Yosef and Roulet, Esteban and Racker, Juan",
    title = "{The Importance of flavor in leptogenesis}",
    eprint = "hep-ph/0601084",
    archivePrefix = "arXiv",
    doi = "10.1088/1126-6708/2006/01/164",
    journal = "JHEP",
    volume = "01",
    pages = "164",
    year = "2006"
}

@article{Abada:2006fw,
    author = "Abada, Asmaa and Davidson, Sacha and Josse-Michaux, Francois-Xavier and Losada, Marta and Riotto, Antonio",
    title = "{Flavor issues in leptogenesis}",
    eprint = "hep-ph/0601083",
    archivePrefix = "arXiv",
    reportNumber = "CERN-PH-TH-2006-001, DNI-UAN-06-03, LPT-ORSAY-06-03, LYCEN-2006-03",
    doi = "10.1088/1475-7516/2006/04/004",
    journal = "JCAP",
    volume = "04",
    pages = "004",
    year = "2006"
}

@article{Abada:2006ea,
    author = "Abada, A. and Davidson, S. and Ibarra, A. and Josse-Michaux, F. -X. and Losada, M. and Riotto, A.",
    title = "{Flavour Matters in Leptogenesis}",
    eprint = "hep-ph/0605281",
    archivePrefix = "arXiv",
    reportNumber = "CERN-PH-TH-2006-093, DNI-UAN-06-97FT, IFT-UAM-CSIC-06-23, LPT-ORSAY-06-21, LYCEN-2006-07",
    doi = "10.1088/1126-6708/2006/09/010",
    journal = "JHEP",
    volume = "09",
    pages = "010",
    year = "2006"
}

@article{Dev:2017trv,
    author = "Dev, P. S. Bhupal and Di Bari, Pasquale and Garbrecht, Bjorn and Lavignac, St{\'e}phane and Millington, Peter and Teresi, Daniele",
    title = "{Flavor effects in leptogenesis}",
    eprint = "1711.02861",
    archivePrefix = "arXiv",
    primaryClass = "hep-ph",
    doi = "10.1142/S0217751X18420010",
    journal = "Int. J. Mod. Phys. A",
    volume = "33",
    pages = "1842001",
    year = "2018"
}

@article{Granelli:2025cho,
    author = "Granelli, Alessandro and Hamaguchi, Koichi and Ramirez-Quezada, Maura E. and Shimada, Kengo and Wada, Juntaro and Yokoyama, Tatsuya",
    title = "{Insights on the scale of leptogenesis from neutrino masses and neutrinoless double-beta decay}",
    eprint = "2502.10093",
    archivePrefix = "arXiv",
    primaryClass = "hep-ph",
    doi = "10.1140/epjc/s10052-025-14487-1",
    journal = "Eur. Phys. J. C",
    volume = "85",
    number = "7",
    pages = "778",
    year = "2025"
}

@article{Davidson:2008bu,
    author = "Davidson, Sacha and Nardi, Enrico and Nir, Yosef",
    title = "{Leptogenesis}",
    eprint = "0802.2962",
    archivePrefix = "arXiv",
    primaryClass = "hep-ph",
    doi = "10.1016/j.physrep.2008.06.002",
    journal = "Phys. Rept.",
    volume = "466",
    pages = "105--177",
    year = "2008"
}

@article{deCesare:2014dga,
    author = "de Cesare, M. and Mavromatos, Nick E. and Sarkar, Sarben",
    title = "{On the possibility of tree-level leptogenesis from Kalb{\textendash}Ramond torsion background}",
    eprint = "1412.7077",
    archivePrefix = "arXiv",
    primaryClass = "hep-ph",
    reportNumber = "KCL-PH-TH-2014-30, LCTS-2014-43",
    doi = "10.1140/epjc/s10052-015-3731-z",
    journal = "Eur. Phys. J. C",
    volume = "75",
    number = "10",
    pages = "514",
    year = "2015"
}

@article{Bossingham:2017gtm,
    author = "Bossingham, Thomas and Mavromatos, Nick E. and Sarkar, Sarben",
    title = "{Leptogenesis from Heavy Right-Handed Neutrinos in CPT Violating Backgrounds}",
    eprint = "1712.03312",
    archivePrefix = "arXiv",
    primaryClass = "hep-ph",
    reportNumber = "KCL-PH-TH-2017-62, IFIC-17-59",
    doi = "10.1140/epjc/s10052-018-5587-5",
    journal = "Eur. Phys. J. C",
    volume = "78",
    number = "2",
    pages = "113",
    year = "2018"
}

@article{Datta:2024xhg,
    author = "Datta, Arghyajit and Manna, Soumen Kumar and Sil, Arunansu",
    title = "{Spontaneous leptogenesis with sub-GeV axionlike particles}",
    eprint = "2405.07003",
    archivePrefix = "arXiv",
    primaryClass = "hep-ph",
    doi = "10.1103/PhysRevD.110.095035",
    journal = "Phys. Rev. D",
    volume = "110",
    number = "9",
    pages = "095035",
    year = "2024"
}

@article{Berbig:2025hlc,
    author = "Berbig, Maximilian",
    title = "{Type II Seesaw Leptogenesis in a Majoron background}",
    eprint = "2506.23290",
    archivePrefix = "arXiv",
    primaryClass = "hep-ph",
    month = "6",
    year = "2025"
}

@article{Chun:2025abp,
    author = "Chun, Eung Jin and Lee, Hyun Min and Song, Jun-Ho",
    title = "{Spontaneous Leptogenesis in Type I Seesaw}",
    eprint = "2512.06413",
    archivePrefix = "arXiv",
    primaryClass = "hep-ph",
    month = "12",
    year = "2025"
}

@article{Kanzaki:2007pd,
    author = "Kanzaki, Toru and Kawasaki, Masahiro and Kohri, Kazunori and Moroi, Takeo",
    title = "{Cosmological Constraints on Neutrino Injection}",
    eprint = "0705.1200",
    archivePrefix = "arXiv",
    primaryClass = "hep-ph",
    reportNumber = "TU-791",
    doi = "10.1103/PhysRevD.76.105017",
    journal = "Phys. Rev. D",
    volume = "76",
    pages = "105017",
    year = "2007"
}

@article{Peng:2025sri,
    author = "Peng, Ying-Quan and Li, Xiu-Fei",
    title = "{Cogenesis of Majoron dark matter and baryon asymmetry from a high scale inverse seesaw mechanism}",
    doi = "10.1140/epjc/s10052-025-14488-0",
    journal = "Eur. Phys. J. C",
    volume = "85",
    number = "7",
    pages = "768",
    year = "2025"
}

@article{Hahn-Woernle:2009jyb,
    author = "Hahn-Woernle, F. and Plumacher, M. and Wong, Y. Y. Y.",
    title = "{Full Boltzmann equations for leptogenesis including scattering}",
    eprint = "0907.0205",
    archivePrefix = "arXiv",
    primaryClass = "hep-ph",
    reportNumber = "CERN-PH-TH-2009-107, MPP-2009-85, PITHA-09-15",
    doi = "10.1088/1475-7516/2009/08/028",
    journal = "JCAP",
    volume = "08",
    pages = "028",
    year = "2009"
}

@article{An:2013yfc,
    author = "An, Haipeng and Pospelov, Maxim and Pradler, Josef",
    title = "{New stellar constraints on dark photons}",
    eprint = "1302.3884",
    archivePrefix = "arXiv",
    primaryClass = "hep-ph",
    reportNumber = "PI-PARTPHYS-318",
    doi = "10.1016/j.physletb.2013.07.008",
    journal = "Phys. Lett. B",
    volume = "725",
    pages = "190--195",
    year = "2013"
}

@article{Redondo:2013lna,
    author = "Redondo, Javier and Raffelt, Georg",
    title = "{Solar constraints on hidden photons re-visited}",
    eprint = "1305.2920",
    archivePrefix = "arXiv",
    primaryClass = "hep-ph",
    reportNumber = "MPP-2013-110",
    doi = "10.1088/1475-7516/2013/08/034",
    journal = "JCAP",
    volume = "08",
    pages = "034",
    year = "2013"
}

@article{Hong:2020bxo,
    author = "Hong, Deog Ki and Shin, Chang Sub and Yun, Seokhoon",
    title = "{Cooling of young neutron stars and dark gauge bosons}",
    eprint = "2012.05427",
    archivePrefix = "arXiv",
    primaryClass = "hep-ph",
    reportNumber = "PNUTP-20/A03, CTPU-20-27, KIAS-P20069",
    doi = "10.1103/PhysRevD.103.123031",
    journal = "Phys. Rev. D",
    volume = "103",
    number = "12",
    pages = "123031",
    year = "2021"
}

@article{Caputo:2021eaa,
    author = "Caputo, Andrea and Millar, Alexander J. and O'Hare, Ciaran A. J. and Vitagliano, Edoardo",
    title = "{Dark photon limits: A handbook}",
    eprint = "2105.04565",
    archivePrefix = "arXiv",
    primaryClass = "hep-ph",
    reportNumber = "NORDITA-2021-036",
    doi = "10.1103/PhysRevD.104.095029",
    journal = "Phys. Rev. D",
    volume = "104",
    number = "9",
    pages = "095029",
    year = "2021"
}

@article{NA64:2023wbi,
    author = "Andreev, Yu. M. and others",
    collaboration = "NA64",
    title = "{Search for Light Dark Matter with NA64 at CERN}",
    eprint = "2307.02404",
    archivePrefix = "arXiv",
    primaryClass = "hep-ex",
    reportNumber = "CERN-EP-2023-130",
    doi = "10.1103/PhysRevLett.131.161801",
    journal = "Phys. Rev. Lett.",
    volume = "131",
    number = "16",
    pages = "161801",
    year = "2023"
}

@article{Bauer:2018onh,
    author = "Bauer, Martin and Foldenauer, Patrick and Jaeckel, Joerg",
    title = "{Hunting All the Hidden Photons}",
    eprint = "1803.05466",
    archivePrefix = "arXiv",
    primaryClass = "hep-ph",
    doi = "10.1007/JHEP07(2018)094",
    journal = "JHEP",
    volume = "07",
    pages = "094",
    year = "2018"
}

@article{delaVega:2023dmw,
    author = "de la Vega, Leon M. G. and Ferro-Hernandez, R. and Garc{\'\i}a-Viltres, A. and Peinado, Eduardo and V{\'a}zquez-J{\'a}uregui, E.",
    title = "{Closing the dark photon window to thermal dark matter}",
    eprint = "2311.17987",
    archivePrefix = "arXiv",
    primaryClass = "hep-ph",
    month = "11",
    year = "2023"
}

@article{Buckley:2011ye,
    author = "Buckley, Matthew R. and Profumo, Stefano",
    title = "{Regenerating a Symmetry in Asymmetric Dark Matter}",
    eprint = "1109.2164",
    archivePrefix = "arXiv",
    primaryClass = "hep-ph",
    reportNumber = "FERMILAB-PUB-11-437-A",
    doi = "10.1103/PhysRevLett.108.011301",
    journal = "Phys. Rev. Lett.",
    volume = "108",
    pages = "011301",
    year = "2012"
}

@article{Tulin:2012re,
    author = "Tulin, Sean and Yu, Hai-Bo and Zurek, Kathryn M.",
    title = "{Oscillating Asymmetric Dark Matter}",
    eprint = "1202.0283",
    archivePrefix = "arXiv",
    primaryClass = "hep-ph",
    reportNumber = "MCTP-12-03",
    doi = "10.1088/1475-7516/2012/05/013",
    journal = "JCAP",
    volume = "05",
    pages = "013",
    year = "2012"
}

@article{Nagata:2016knk,
    author = "Nagata, Natsumi and Olive, Keith A. and Zheng, Jiaming",
    title = "{Asymmetric Dark Matter Models in SO(10)}",
    eprint = "1611.04693",
    archivePrefix = "arXiv",
    primaryClass = "hep-ph",
    reportNumber = "UT--16--31, UMN--TH--3612-16, FTPI--MINN--16-32",
    doi = "10.1088/1475-7516/2017/02/016",
    journal = "JCAP",
    volume = "02",
    pages = "016",
    year = "2017"
}

@article{Ghosh:2021qbo,
    author = "Ghosh, Avirup and Ghosh, Deep and Mukhopadhyay, Satyanarayan",
    title = "{Cosmology of complex scalar dark matter: Interplay of self-scattering and annihilation}",
    eprint = "2103.14009",
    archivePrefix = "arXiv",
    primaryClass = "hep-ph",
    doi = "10.1103/PhysRevD.104.123543",
    journal = "Phys. Rev. D",
    volume = "104",
    number = "12",
    pages = "123543",
    year = "2021"
}

@article{PandaX-4T:2021bab,
    author = "Meng, Yue and others",
    collaboration = "PandaX-4T",
    title = "{Dark Matter Search Results from the PandaX-4T Commissioning Run}",
    eprint = "2107.13438",
    archivePrefix = "arXiv",
    primaryClass = "hep-ex",
    doi = "10.1103/PhysRevLett.127.261802",
    journal = "Phys. Rev. Lett.",
    volume = "127",
    number = "26",
    pages = "261802",
    year = "2021"
}

@article{XENON:2018voc,
    author = "Aprile, E. and others",
    collaboration = "XENON",
    title = "{Dark Matter Search Results from a One Ton-Year Exposure of XENON1T}",
    eprint = "1805.12562",
    archivePrefix = "arXiv",
    primaryClass = "astro-ph.CO",
    doi = "10.1103/PhysRevLett.121.111302",
    journal = "Phys. Rev. Lett.",
    volume = "121",
    number = "11",
    pages = "111302",
    year = "2018"
}

@article{CRESST:2024cpr,
    author = "Angloher, G. and others",
    collaboration = "CRESST",
    title = "{First observation of single photons in a CRESST detector and new dark matter exclusion limits}",
    eprint = "2405.06527",
    archivePrefix = "arXiv",
    primaryClass = "astro-ph.CO",
    doi = "10.1103/PhysRevD.110.083038",
    journal = "Phys. Rev. D",
    volume = "110",
    number = "8",
    pages = "083038",
    year = "2024"
}

@article{CRESST:2019jnq,
    author = "Abdelhameed, A. H. and others",
    collaboration = "CRESST",
    title = "{First results from the CRESST-III low-mass dark matter program}",
    eprint = "1904.00498",
    archivePrefix = "arXiv",
    primaryClass = "astro-ph.CO",
    doi = "10.1103/PhysRevD.100.102002",
    journal = "Phys. Rev. D",
    volume = "100",
    number = "10",
    pages = "102002",
    year = "2019"
}

@article{PICO:2019vsc,
    author = "Amole, C. and others",
    collaboration = "PICO",
    title = "{Dark Matter Search Results from the Complete Exposure of the PICO-60 C$_3$F$_8$ Bubble Chamber}",
    eprint = "1902.04031",
    archivePrefix = "arXiv",
    primaryClass = "astro-ph.CO",
    reportNumber = "FERMILAB-PUB-19-073-AE-E",
    doi = "10.1103/PhysRevD.100.022001",
    journal = "Phys. Rev. D",
    volume = "100",
    number = "2",
    pages = "022001",
    year = "2019"
}

@article{PICO:2015pux,
    author = "Amole, C. and others",
    collaboration = "PICO",
    title = "{Dark matter search results from the PICO-60 CF$_3$I bubble chamber}",
    eprint = "1510.07754",
    archivePrefix = "arXiv",
    primaryClass = "hep-ex",
    reportNumber = "FERMILAB-PUB-15-449-AE-E",
    doi = "10.1103/PhysRevD.93.052014",
    journal = "Phys. Rev. D",
    volume = "93",
    number = "5",
    pages = "052014",
    year = "2016"
}

@article{LZ:2024zvo,
    author = "Aalbers, J. and others",
    collaboration = "LZ",
    title = "{Dark Matter Search Results from 4.2{\,}{\,}Tonne-Years of Exposure of the LUX-ZEPLIN (LZ) Experiment}",
    eprint = "2410.17036",
    archivePrefix = "arXiv",
    primaryClass = "hep-ex",
    reportNumber = "FERMILAB-PUB-24-0796-V",
    doi = "10.1103/4dyc-z8zf",
    journal = "Phys. Rev. Lett.",
    volume = "135",
    number = "1",
    pages = "011802",
    year = "2025"
}

@article{Mojahed:2025exj,
    author = "Mojahed, Martin A. and Weber, Sascha",
    title = "{Asymgenesis}",
    eprint = "2507.10655",
    archivePrefix = "arXiv",
    primaryClass = "hep-ph",
    reportNumber = "MITP-25-051",
    doi = "10.1016/j.physletb.2026.140294",
    journal = "Phys. Lett. B",
    volume = "875",
    pages = "140294",
    year = "2026"
}

@article{Abhishek:2026hex,
    author = "Abhishek and Mummidi, V. Suryanarayana",
    title = "{Asymmetric dark matter from leptogenesis in type-III seesaw framework with modular $S_4$ symmetry}",
    eprint = "2602.03384",
    archivePrefix = "arXiv",
    primaryClass = "hep-ph",
    month = "2",
    year = "2026"
}

@article{McKenna:2026xar,
    author = "McKenna, Henry G. F. and Smirnov, Juri and Gorbahn, Martin",
    title = "{Linking Leptogenesis and Asymmetric Dark Matter: A Testable Framework for Neutrino Mass and the Matter-Antimatter Asymmetry}",
    eprint = "2602.20237",
    archivePrefix = "arXiv",
    primaryClass = "hep-ph",
    month = "2",
    year = "2026"
}

@article{Borah:2026vqw,
    author = "Borah, Debasish and Paul, Partha Kumar and Sahu, Narendra",
    title = "{Cogenesis of visible and dark matter in type-I Dirac seesaw}",
    eprint = "2603.24693",
    archivePrefix = "arXiv",
    primaryClass = "hep-ph",
    month = "3",
    year = "2026"
}

\end{document}